\documentclass[twocolumn,showpacs,preprintnumbers,amsmath,amssymb]{revtex4}
\usepackage{axodraw}
\usepackage{graphicx}% Include figure files
\newcommand{\be}{\begin{equation}}  
\newcommand{\ee}{\end{equation}}  
\newcommand{\bear}{\begin{eqnarray}}  
\newcommand{\eear}{\end{eqnarray}}  
\newcommand{\ba}{\begin{array}}  
\newcommand{\ea}{\end{array}}

\newskip\humongous \humongous=0pt plus 1000pt minus 1000pt

\newif\ifdtup

\def\oldreffmt#1{\rlap{[#1]} \hbox to 2\parindent{}}

\def\figfmt#1{\rlap{Figure {#1}} \hbox to 1in{}}  
  
\def\ie{\hbox{\it i.e.}{}}	\def\etc{\hbox{\it etc.}{}}  
\def\eg{\hbox{\it e.g.}{}}

\def\VEV#1{\left\langle #1\right\rangle}

\def\beq{\begin{equation}}  
\def\eeq{\end{equation}}  
\def\bea{\begin{eqnarray}}  
\def\eea{\end{eqnarray}}  
\def\half{\frac{1}{2}}  
  
\def\bq{\begin{quote}}  
\def\eq{\end{quote}}

\def\GeV{\,{\rm GeV}}

\def\half{\frac{1}{2}}       
\def \lta {\mathrel{\vcenter  
     {\hbox{$<$}\nointerlineskip\hbox{$\sim$}}}}  
   
\relax  

\newdimen\tdim  
\tdim=\unitlength  
\def\bar{\overline}

\begin{document}

\preprint{FERMILAB-Pub-14-003-T}
\title{
Is the Higgs Boson Associated with
\\ Coleman-Weinberg Dynamical Symmetry Breaking?}

\author{$ $ \\
Christopher T. Hill
%\\ and \\
%Graham G. Ross$^2$
\\ $ $}

%\email{hill@fnal.gov}

\affiliation{
 {
{Fermi National Accelerator Laboratory}}\\
{{\it P.O. Box 500, Batavia, Illinois 60510, USA}}
%\\  $ $ \\
%$^2$Department of Theoretical Physics,\\
%University of Oxford,1 Keble Road,\\
%Oxford OX1 3NP
}

\date{\today}% It is always \today, today,

\begin{abstract}
The Higgs mechanism may be a quantum phenomenon, \ie, a Coleman-Weinberg 
potential generated by the explicit breaking of scale symmetry in
Feynman loops.  We review the relationship of scale symmetry, 
trace anomalies, and emphasize the role of the renormalization group 
in determining Coleman-Weinberg potentials.  We propose a simple phenomenological model with
``maximal visibility'' at the LHC containing a ``dormant''  Higgs doublet (no VEV,
coupled  to  standard model gauge interactions $SU(2)\times U(1)$) with a  
mass of $\sim 380$ GeV.   We  discuss
the LHC phenomenology and UV challenges of such a model.
We also give a schematic model in which  new heavy fermions,
with  masses $\sim 230$ GeV, 
can drive a Coleman-Weinberg potential at two-loops.  The role of the
``improved stress tensor''  is emphasized, 
and we propose a non-gravitational term, analogous
to the $\theta$-term in QCD, which generates it from a scalar action.
\end{abstract}

\maketitle

\noindent
{\bf I. Introduction}
\vskip .1in
\noindent

The discovery of a light Higgs boson 
presents several well-known puzzles: 
``What mechanism determines the mass of the Higgs boson?" ``What 
is the custodial symmetry that yields a small mass scale for the
apparently pointlike $0^+$ particle?"  ``Is there any
new associated dynamics with the Higgs boson?"
These questions revolve around the problem of the naturalness of the existence
of a low mass fundamental spin-$0$ field in quantum field theory.

Supersymmetry offers 
solutions to these problems.  The custodial symmetry of light
Higgs bosons could be the chiral symmetry of their superpartners, \eg, the
fermionic Higgsinos.  This, however,
requires proximity in mass scale of SUSY states and, so far, supersymmetry has not emerged in searches.
SUSY is also highly constrained by the relatively heavy $\sim 125$ GeV Higgs
boson mass, and indirect measures, 
such as the electron EDM and $b\rightarrow s\gamma$, \etc.
While SUSY remains a popular candidate for an ultimate
solution to these problems,  its relevance to
the electroweak scale has become somewhat clouded by the necessity
of a high degree of fine-tuning \cite{S1,S2,S3,S4}. SUSY as the
custodial symmetry of the Higgs boson will be subject
to more definitive tests in Run-II of the LHC, circa 2015-18. 

Strong dynamics also offers 
natural solutions, by postulating a 
mechanism similar to that of QCD for
electroweak symmetry breaking (see the review, \cite{simmons}). QCD  involves explicit 
breaking of scale symmetry
via the trace anomaly, proportional to the $\beta$-function of the QCD coupling constant.
The QCD mass scale, which accounts for the masses
of the nucleons and, hence, most of the visible mass in the universe,
 is an example of mass generation
by ``dimensional transmutation.'' This is inherently
a quantum phenomenon, \ie, it is {\em  mass generated from quantum mechanics itself}. 
The QCD hierarchy arises naturally since the 
ratio of the QCD scale $\Lambda_{QCD}$,
to any large scale in nature, $M$, is
given:
\beq
\label{one}
\frac{\Lambda_{QCD}}{M} = \exp\left(-\frac{8\pi^2}{|b_0|g^2(M)}   \right)
\eeq
where $b_0 = [\hbar ](11 -(2/3)n_f)$ at one-loop precision. 
Here one inputs
a small dimensionless coupling constant, $g^2(M)$, at
an arbitrary high energy scale, $M$. Quantum loops
then generate the small ratio, $\Lambda_{QCD}/M$.
Eq.(\ref{one})
implies that the 't Hooft naturalness, \ie, the custodial
symmetry associated with the smallness of
the ratio,  ${\Lambda_{QCD}}/{M}\rightarrow 0$, which occurs when
  $b_0\rightarrow 0$, 
has the interpretation of
 the {\em classical scale symmetry} 
of QCD in the $\hbar \rightarrow 0$ limit  \cite{bardeen2,Hill10}. 

Straightforward attempts to implement
an analogous QCD-like mechanism for generating
the weak scale have generally failed. This
approach
typically yields light $0^-$ boundstates,  \eg, Nambu-Goldstone techni-pions. 
These couple perturbatively to $ZZ$ and $WW$, through axial anomalies,
and therefore cannot be imposters of a Higgs boson which couples
at tree level and is consistent with present experimental indications. 
There is effort underway to construct viable scenarios
(for a partial list, see \eg, refs.\cite{Appelquist, Shrock, Miransky, BHL, Fukano, Dobrescu}), 
but strong dynamical models,
as a class, have been even more severely constrained by LHC data than SUSY.   

The present evidence from the LHC
strongly favors a simple perturbative Higgs 
boson interpretation of the data as proposed by
Weinberg in 1967 \cite{Weinberg}.  But to date
we have no understanding as to the origin of the electroweak scale,
first introduced by Fermi in 1934 \cite{Fermi}. 

Presently
we wish to focus upon an alternative approach. We will
argue for  a quantum origin of the Higgs potential
and electroweak scale: We propose that the Higgs potential
is a perturbative Coleman-Weinberg potential \cite{Coleman:1973jx}. 
As such, we ask what the current  data might be telling us and what might
be visible consequences of this hypothesis at the LHC in Run-II and beyond \cite{CWCTH}.
We emphasize at the outset that we will not delve in
great detail into the UV completion
aspects of this idea. 
We think the question, ``Is the Higgs potential generated by quantum mechanics?''
to be sufficiently compelling that it should be posed in a self-contained framework,
and addressed experimentally in Run-II and beyond.

Coleman-Weinberg (CW) symmetry breaking is  complimentary
to a QCD-like, strong dynamical mechanism.  It 
arises from a stress-tensor trace anomaly, \ie, 
it relies upon scale symmetry breaking by perturbative quantum loops. 
This means that CW symmetry breaking can be understood entirely in terms of 
the renormalization group (RG) running of the the Higgs scalar quartic coupling constant,
in analogy to QCD.  We discuss this in greater formal detail in Section II and Appendices A-C
and we'll introduce a few new ideas.  

With a CW potential  the custodial symmetry
for the weak scale again arises like QCD,  as the scale invariance
of the action in the $\hbar \rightarrow 0$ limit. In this limit quantum loops 
are turned off and the trace anomaly goes to zero.  The ``improved stress tensor,'' \cite{CCJ},
defines the renormalization group of the CW potential, and the trace anomaly is
determined as $-(\beta/\lambda)V(\phi)$. We see that $\beta/\lambda $ is
the anomalous dimension of the potential. CW symmetry breaking occurs
at a local minimum of the potential, where the 
anomalous dimension takes on the value $-4$ and the $d=4$ potential
operator becomes pure $d=0$ vacuum energy.  

The CW
potential, expanded about its minimum, depends only upon
the local values of RG $\beta$-functions and their derivatives.
We give an  expression valid to all orders in perturbation theory, 
through quintic order in the Higgs field, for the CW potential.  Incidently, in Appendix A 
we introduce a novel, non-gravitational term, into the scalar field action 
that is a non-topological analogue of the $\theta$-term in QCD, but which generates the improved
stress tensor from variation of the action.

The idea that classical scale symmetry can arguably serve as
a custodial symmetry of a fundamental perturbative Higgs boson has been
emphasized by Bardeen \cite{bardeen, bardeen2}. 
 In implementation of the CW mechanism to
obtain the observed value of  $v_{weak} = 175$ GeV and Higgs boson mass,
$m_h = 125$ GeV, we find, in its simplest and most obvious incarnation,  that
 additional large bosonic contributions to the RG equation
for the Higgs quartic coupling are required.  Recently, various authors have focused on
related models, many of which accomplish this with bosonic dark matter fields,  
\cite{pilaf1,Hambye1,Hambye2,Dermisek,general}. We will presently examine
a ``maximally visible new physics scenario at LHC'' to implement
the CW mechanism.  We will also propose a novel
mechanism for generating a CW potential from fermions 
that emerges upon a more detailed scrutiny of the RG (see Section V).

%\newpage
In Sections III and IV we consider a bosonic model consisting of a second, 
``dormant,'' Higgs boson as the source of new bosonic contributions
to the RG equations to sculpt the CW potential.  
Here we distinguish the oft-used term ``inert doublet,'' to imply that the second
Higgs doublet {\em does not interact} with standard model $SU(2)\times U(1)$ gauge fields, (\eg, as
in the dark matter models of \cite{Hambye1,Hambye2,Dermisek,general}), from the term
``{\em dormant} Higgs doublet.''  By ``dormant'' we imply that the second doublet does interact
with standard $SU(2)\times U(1)$ interactions, however the second doublet has no VEV. 
Such dormant Higgs doublet models are valid
solutions to the original Weiberg-Glashow natural, \cite{WeinbergGlashow},
two-Higgs doublet schemes \cite{CTH, Deshpande}.
Our question is: ``How accessible is the 
dormant Higgs doublet at the LHC?''  We 
estimate production and decay rates 
and, modulo a more thorough LHC detector based analysis, the
results are encouraging.

Phenomenologically, we find that the new  dormant Higgs doublet  must have a mass
of about $\sim 380$ GeV. Since we assume standard model
couplings, it is  guaranteed to be pair-produced, above threshold of $\sim 800$ GeV, via
$q\bar{q}\rightarrow (\gamma^*, Z^*, h)\rightarrow (H^0 H^{0\dagger}, H^+H^-)$ 
and $q\bar{q}\rightarrow W^*\rightarrow H^+H^0$ at the LHC.
Other  production and decay channels
are likely, but model dependent.  We think it is most  natural,
albeit an additional assumption, that the  dormant doublet couples to 
 $b$ quarks,  $\sim (\bar{t}, \bar{b})_LH'b_R$ with a large
$O(1)$ coupling constant, $g_b'$.   This makes the
dormant doublet the natural flavor partner of the Higgs with
it's large coupling to the top quark.
These $b$-quark couplings allow enhanced production
of single $H^0$ and $H^\pm$ in association with $b\bar{b}$
and $\bar{t}b$ or $t\bar{b}$, and  would also imply decays like $ H^0\rightarrow b\bar{b}$ and
$ H^+\rightarrow t\bar{b}$ which become interesting observables at the LHC.
 
One intriguing corollary associated with the
CW potential is that the Higgs potential
will have cubic, quadrilinear, even quintic (and
higher order) coupling constants, that will be significantly 
different than those of 
the standard model \cite{CWCTH, Dermisek}.
 
In Section V we also present a schematic model of a Coleman-Weinberg potential
for the Higgs  generated by new fermions. This is a novel
approach, and arises from a two-loop effect in the RG structure
of the CW potential.  We think this class of models may alleviate
some of the potential problems encountered with new heavy bosons in UV completion, 
which requires further  development \cite{Ross}. While we have not examined the full UV structure
or phenomenological implications of this scheme, it suggests pair produced new fermions
with masses $\sim 200$ GeV.  These fermions would have their own strong interaction,
and may be produced in boundstates 
with a threshold at $\sim 400$ GeV, 
or pairs of new heavy meson-like boundstates at $\sim 800$ GeV.  

We begin by discussing some general theoretical aspects of Coleman-Weinberg symmetry
breaking. 

\newpage
\vskip .3in
\noindent
{\bf II. General Theoretical Considerations}
\vskip .1in

\subsection{Schematic Analysis}

To get a feeling for how Coleman-Weinberg
symmetry breaking works, with particular emphasis upon
the renormalization group (RG), we consider
a $U(1)$ Higgs scalar field potential 
$ \half \lambda (H^\dagger H)^2$. This classical
potential, for $\lambda > 0$  has an uninteresting minimum at $\VEV{H}= v =0$
as in Fig.(1).

A process with $N$ Feynman loops is of order $O(\hbar^N)$
in field theory.  Quantum loops lead to the RG
 running of couplings, such as $\lambda$, with scale, $\mu$.
Typically, we might have a one-loop, $O(\hbar)$, solution to the RG equations
as in Fig.(2):
\beq
\lambda(\mu)
\approx  \beta \ln(\mu/M)
\eeq
where $\beta \propto \hbar$.  
$M$ simply parameterizes  the
particular RG trajectory of the running  $\lambda(\mu) $, \ie, 
we would ask our experimental colleagues to 
measure the dimensionless quantity
 $\lambda$ at some energy scale $\mu$, and we would then choose 
$M$ so that we fit their result as
$\lambda_{expt}(\mu) = \beta \ln(\mu /M)$.
 
%\begin{figure}[h]
%\includegraphics[bb = 0 0 100 100 ]{UV1.pdf}
%\caption{UV running of the Dormant Higgs model}
%\end{figure}
\begin{figure}[t]
\vspace{5.5cm}
\includegraphics{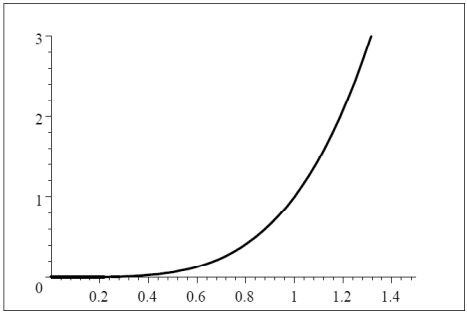}
\vspace{-0.0cm}
\caption[]{ Classical $\sim \lambda v^4 $ potential. }
\label{xroots}
\end{figure}

%\begin{figure}[h]
%\includegraphics[bb = 0 0 100 100 ]{UV1.pdf}
%\caption{UV running of the Dormant Higgs model}
%\end{figure}
\begin{figure}[t]
\vspace{5.5cm}
\includegraphics{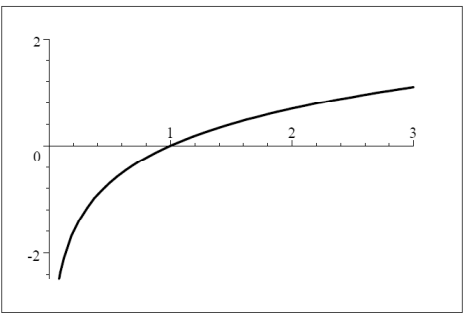}
\vspace{-0.4cm}
\caption[]{ Typical RG trajectory $\lambda \sim  \beta \ln(v/M)$}
\label{xroots}
\end{figure}

%\begin{figure}[h]
%\includegraphics[bb = 0 0 100 100 ]{UV1.pdf}
%\caption{UV running of the Dormant Higgs model}
%\end{figure}
\begin{figure}[t]
\vspace{5.5cm}
\includegraphics{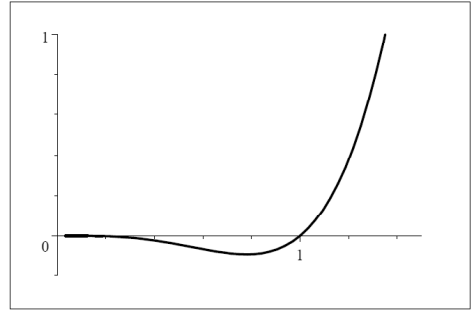}
\vspace{-0.0cm}
\caption[]{ Resulting CW potential, $\sim  \beta v^4 \ln(v/M)$}
\label{xroots}
\end{figure}

With the running quartic coupling constant, the scale can be set by
the vacuum expectation value,  $\VEV{H} = v$,  of the field $H$ itself. 
The resulting scalar potential, as a function
of  $v= \mu $, is then:
\beq
\label{pot}
V(v)=  \half \beta v^4 \ln(v/M)
\eeq
This potential has
a local minimum, as in Fig.(3),  at   $v_0 = Me^{-1/4} $.
A stable minimum of the potential, \ie, one with a positive
curvature at the minimum, or $m_h^2 > 0$,  occurs just below the zero-crossing of 
$\lambda(v)$ from a negative to a positive value.  Hence $\lambda$
must be negative  and $\beta$ must be positive
at the minimum (we'll see in Section V that there are alternative
solutions involving two loops in  which  the situation is flipped, \ie,
$\beta$ ($\lambda$) can be negative (positive)).

If $\lambda(\mu) $ continues to run as  $\propto\beta \ln(\mu/M)$,
we would see that the ratio of the VEV, $v_0$, to any other scale $M'$ is then:
\beq
\frac{v_{0}}{M'} \propto \exp\left(-\frac{\lambda(M')}{ \beta}\right)
\eeq
 A large hierarchy between $v_0$ and $M'$ 
can be exponentially controlled by the ratio of dimensionless quantities, $\lambda(M')/\beta$.
With $v_{weak}\equiv v_0$,
the  't Hooft naturalness of the ``small ratio'' of ${v_{weak}}/{M_{Planck}}$, or
${v_{weak}}/{M_{GUT}}$, would be, in analogy to QCD,  associated with the 
limit $\beta\rightarrow 0$,  which is
again the limit of classical scale invariance, $\hbar\rightarrow 0$.
Of course, the RG running of $\lambda(\mu) $ can be  complicated
over a large range of $\mu$. 

$\beta$, which we have approximated as a constant above,
 is the  $\beta$-function of $\lambda$ which defines the
Gell-Mann--Low renormalization group equation \cite{MGM}: 
\beq
\label{RG0}
\frac{d\lambda}{d\ln(\mu)} = \beta(\lambda) 
\eeq

To see the structure of the CW potential
in somewhat greater detail we expand the potential of eq.(\ref{pot}) in $H$ about
a hypothetical vacuum expectation value $v$:
\beq
|{H}| =v+h/\sqrt{2}
\eeq
where $h$ is a physical Higgs boson field,
according to $\lambda(|{H}|)
=  \beta \ln(|{H}|/M)$:
\bea
\lambda (v+h/\sqrt{2})  & = & \lambda (v)+\beta\ln (1+h/\sqrt{2}v )
\nonumber \\ 
& & \!\!\!\!\!\!\!\!\!  \!\!\!\!\!\!\!\!\!  \!\!\!\!\!\!\!\!\!  
\!\!\!\!\!\!\!\!\!  \!\!\!\!\!\!\!\!\!
\approx  \lambda(v) + \beta\left( \frac{h}{\sqrt{2}v}-\frac{h^2}{4v^{2}}
+\frac{h^{3}}{6\sqrt{2} v^{3}}-\frac{h^{4}}{16 v^{4}}+\frac{h^{5}}{20\sqrt{2} v^{5}}
 \right) 
\nonumber \\
& & +{\cal{O}}\left( h^{6}\right) 
 \eea
We thus have the CW Higgs potential:
\bea 
\label{CW1}
V_{CW}(h)   & = & \half \lambda (v+h/\sqrt{2})(v+h/\sqrt{2})^{4}
\nonumber \\
 & &\!\!\!\!\!\!\!\!\!\!    \!\!\!\!\!\!\!\!\!\!   \!\!\!\!\!\!\!\!\!\!    \!\!\!\!\!\!\!\!\!\! 
 =  \half \lambda v^{4} +\left( \lambda +\frac{1}{4}\beta \right) 
\sqrt{2}v^{3}h+\left( \frac{3}{2}\lambda +\frac{7}{8}\beta \right)
v^{2}h^{2}\allowbreak 
\nonumber \\
 & & \!\!\!\!\!\!\!\!\!\!    \!\!\!\!\!\!\!\!\!\!   \!\!\!\!\!\!\!\!\!\!     
+ \left(\half \lambda +\frac{13}{24}\beta \right) v
\sqrt{2}h^{3}+\left( \frac{1}{8}\lambda +\frac{25}{96}\beta \right) h^{4} 
\nonumber \\
& & \!\!\!\!\!\!\!\!\!\!    \!\!\!\!\!\!\!\!\!\!   \!\!\!\!\!\!\!\!\!\!
+\frac{1}{40\sqrt{2}v}\beta  h^{5}  +  O\left( h^{6}\right) 
\eea
The extremum of the potential is given by:
\beq
\left. \frac{dV}{dh}\right|_{h=0} = \sqrt{2}v^{3}\left( \lambda +\frac{1}{4}\beta \right) =0
\eeq
which requires:
\beq
\label{oneten0}
\beta =-4\lambda .
\eeq
The Higgs boson mass is then given by: 
\beq
\frac{d^2V}{dh^2} = m_{h}^{2}=\left( 3\lambda +\frac{7}{4}\beta \right) v^{2} 
\eeq
If the extremum is to be a minimum
we must impose the positivity condition:
\beq
m_{h}^{2} > 0
\eeq 
Therefore, from equation (\ref{oneten0}):
\beq
\label{twentyseven0}
 m_{h}^{2}=-4\lambda v^2 = \beta v^{2} >0
\eeq
This shows that $\beta > 0$ and $\lambda < 0$ at the
minimum of the potential (we'll see in Section V
that there is a flipped solution arising at two-loops with $\lambda > 0$
and $\beta < 0$).

The resulting Higgs potential, expanded about the minimum and using eq.(\ref{oneten0})
 through quintic order, is: 
\bea 
V_{CW}(h) &  = & -\frac{1}{8}\beta v^{4} + \frac{1}{2} \beta v^{2}h^{2} 
+\frac{5}{6\sqrt{2}}\beta  vh^{3}+\frac{11}{48}\beta h^{4}
\nonumber \\
& & +\frac{1}{40\sqrt{2}v}\beta  h^{5}  +  O\left( h^{6}\right) 
\eea

\subsection{All-Orders RG Improved Potentials}

The above analysis of the CW Higgs potential is schematic, relying upon a particular solution
and expanding in a leading logarithm. In Section II.C we'll see that the trace anomaly of the
``improved stress tensor'' implies an exact equation for the Coleman-Weinberg potential of a
scalar field $\phi$:
\beq
\label{CWeqn}
\phi\frac{\delta}{\delta\phi}V(\phi) - 4V(\phi) = \frac{\beta}{\lambda}V(\phi)
\eeq
We can view this as the definition of the CW potential for  the VEV of $\phi$. Here $\beta$ is
the all-orders $\beta$ function of $\lambda$. 
The solution is:
\beq
V(\phi) =\half \lambda(\phi) \phi^4\qquad \makebox{where}\qquad \frac{d\lambda(\mu)}{d\ln\mu} =\beta(\lambda)
\eeq
(the normalization $\half$ is locked to the definition of the classical potential, and defines $\beta(\lambda) $).
This is formal, but
a useful application to the previous model of Section II.A, with $\phi\sim |H|$,
applies to both CW and standard model Higgs
potentials, involves
the expansion of $\lambda(v+h/\sqrt{2})$ in terms of
the  $\beta$-functions computed to all-orders of perturbation
theory in all relevant couplings.  

We label all relevant coupling constants that  enter in any order
of the loop diagrams for the running of  $\lambda$ (\eg, $g_{top}$, $g_2$, $g_{QCD}$, etc.)
 as $\lambda_i$.   We denote  the scalar quartic 
(Higgs) coupling as $\lambda\equiv \lambda_1$
with $\beta$-function $\beta_1(\lambda_i)$.  Each $\lambda_i$  has
its own $\beta_i$:
\beq
\frac{d\lambda_i}{d\ln(\mu)} = \beta_i(\lambda_j)
\eeq
The derivatives of $\lambda_1$  about
the VEV $v$ can be written in terms of
the $\beta$-functions:
\bea
v\lambda'_1(v) &=& \beta_1
\\
v^2\lambda''_1(v) & = &  \beta_j\frac{\partial \beta_1}{\partial\lambda_j} -\beta_1
\\
v^3\lambda'''_1(v) & = & 
 \beta_i\beta_j \frac{\partial^2 \beta_1}{\partial\lambda_i \partial\lambda_j} 
+ \beta_j\frac{\partial \beta_i}{\partial\lambda_j} \frac{\partial \beta_1}{\partial\lambda_i } 
\nonumber \\
& & 
-3 \beta_j\frac{\partial \beta_i}{\partial\lambda_i}+2\beta_1 
\\
v^4\lambda''''_1(v) & = & 
\beta_i\beta_j \beta_k \frac{\partial^3 \beta_1}{\partial\lambda_i \partial\lambda_j\partial\lambda_k} 
+\beta_k\frac{\partial \beta_j}{\partial\lambda_k} \frac{\partial \beta_i}{\partial\lambda_j }\frac{\partial \beta_1}{\partial\lambda_i }
 \nonumber \\
& & 
+ \beta_k \beta_j\frac{\partial^2 \beta_i}{\partial\lambda_j\partial\lambda_k} \frac{\partial \beta_1}{\partial\lambda_i } 
+ 3\beta_k \beta_j\frac{\partial \beta_i}{\partial\lambda_k} \frac{\partial^2 \beta_1}{\partial\lambda_i \partial\lambda_j }
\nonumber \\
& & 
 -6 \beta_i \beta_j\frac{\partial^2 \beta_1}{\partial\lambda_i \partial\lambda_j } 
- 6 \beta_k \frac{\partial \beta_j}{\partial\lambda_k }\frac{\partial \beta_1}{\partial\lambda_j } 
\nonumber \\
& &
 + 11\beta_i\frac{\partial \beta_1}{\partial\lambda_i}-6\beta_1
%\nonumber \\
%\!\!\!\!\!  \!\!\!\!\!  \!\!\!\!\!  \!\!\!\!\!  \!\!\!\!\!   
% \!\!\!\!\!  \!\!\!\!\!  \!\!\!\!\!  \!\!\!\!\!
\eea
(we tabulate the quintic order in Appendix D).  
Each $\beta_i=\beta_i(\lambda_j(v))$ is a function
of the couplings, $ \lambda_j(v) $, evaluated at the scale, $v$. We use
the summation convention for repeated indices, $i,j,k$.
%Note that these relations are recursive:
%\beq
%v^N\lambda^{[N]}_1(v)  = v\frac{d}{dv} (v^{N-1}\lambda^{[N-1]})-(N-1)v^{N-1}\lambda^{[N-1]} .
%\eeq

The leading terms in the CW potential now take the form to all orders
in $\hbar$:
\bea 
V_{CW}(h) &  = & \frac{1}{2}\lambda(v) v^{4} 
+\sqrt{2}\left( \lambda_1 +\frac{1}{4}\beta_1 \right) 
v^{3}h
\nonumber \\
& & +\; O(h^3)  \eea
The extremum condition
is therefore formally the same as in the schematic case:
\beq
\label{extremum10}
\beta_1(\lambda_i(v)) = -4 \lambda_1(v)
\eeq
but note that this is now an all-orders in $\hbar$ condition
and, likewise, the anomalous dimension of the potential
when we impose the extremum is $\beta_1/\lambda_1 =-4$
to all orders.

Imposing the extremum condition, eq.(\ref{extremum10}), we obtain the Coleman-Weinberg 
potential expanded about the 
energy minimum:
\bea 
\label{CWfull}
V_{CW}(h) &  = & -\frac{1}{8}\beta_1 v^{4}  
 + \frac{1}{2}v^{2}h^{2}  \left( \beta_1  + \frac{1}{4}\beta_j\frac{\partial \beta_1}{\partial\lambda_j} \right)
\nonumber \\
 &  & \!\!\!\!\!\! 
+ \frac{5}{6\sqrt{2}} vh^{3}\left( \beta_1 
+\frac{9}{20}\beta _{i}\frac{\partial \beta_1 }{\partial \lambda_{i}}
+ \frac{1}{20}\beta_{j}\beta _{i}\frac{\partial ^{2}\beta_1 }{\partial \lambda_{j}\partial \lambda_{i}}\right.
\nonumber \\
& &
\left.
+ \frac{1}{20}\beta _{j}\frac{\partial \beta _{i}}{\partial \lambda_{j}}\frac{\partial \beta_1 }{\partial\lambda_{i}}\right) 
\nonumber \\
& & \!\!\!\!\!\! 
+
\frac{11}{48} h^{4}\left( 
\beta_1 
+\frac{35}{44}  \beta_{i}\frac{\partial \beta_1 }{\partial \lambda_{i}} 
+ \frac{5}{22}\beta_{j}\beta_{i}\frac{d^{2}\beta_1 }{\partial \lambda_{j}\partial \lambda_{i}}
 \right.
\nonumber \\
& & + \frac{5}{22} \beta_{j}\frac{\partial \beta_{i}}{\partial \lambda_{j}}\frac{\partial \beta_1 }{\partial \lambda_{i}} 
+ \frac{1}{44}\beta_{k}\beta_{j}\beta_{i}\frac{d^{3}\beta_1 }{\partial \lambda
_{k}\partial \lambda_{j}\partial \lambda_{i}}
\nonumber \\
& &
\left. +
 \frac{1}{44}\beta_{k}\frac{\partial \beta_{j}}{\partial \lambda_{k}}
\frac{\partial \beta_{i}}{\partial \lambda_{j}}\frac{\partial \beta_1 }{\partial \lambda_{i}} 
+ \frac{1}{44}\beta_{j}\beta_{i}\frac{d^{2}\beta_{i}}{\partial \lambda_{j}\partial \lambda_{i}}\frac{\partial \beta_1 
}{\partial \lambda_{i}}\right.
\nonumber \\
& & \left. 
+\frac{3}{44}\beta_{j}\beta_{k}\frac{\partial \beta_{i}}{\partial \lambda_{k}}
\frac{d^{2}\beta_1 }{\partial \lambda_{j}\partial \lambda_{i}}  
\right) +  ...
\eea 
(the quintic term, $...$, is tabulated in Appendix D).
As a check on these results, notice that if we keep
only the leading $O(\hbar)$ terms,  
we recover the schematic model case of eq.(\ref{CW1}).

Note that we can also apply this expansion to the standard model:
\bea
\label{SMpot}
V_{SM} & = & \half\tilde{\lambda}(v+h/\sqrt{2})\left(v^2 - (v+h/\sqrt{2})^2\right)^2
\nonumber \\
& = & 
\tilde\lambda v^{2}h^{2}
+\frac{1}{\sqrt{2}}\left( \tilde\lambda +\tilde\beta_1 \right) vh^{3}
\nonumber \\
& & \!\!\!\!\!
+\frac{1}{4}\left( \frac{1}{2}\tilde\lambda +\tilde\beta
+ \tilde\beta _{i}\frac{\partial\tilde\beta_1 }{\partial\tilde\lambda _{i}}  \right) h^{4}
\nonumber \\
& & 
%\!\!\!\!\!   \!\!\!\!\! \!\!\!\!\!  \!\!\!\!\! 
\!\!\!\!\! 
 + \frac{1}{24\sqrt{2}}\left( \tilde\beta +2\tilde\beta _{j}\tilde\beta _{i}\frac{\partial^{2}\tilde\beta_1 }{
\partial\tilde\lambda _{j}\partial\tilde\lambda _{i}}+2\tilde\beta _{j}\frac{\partial\tilde\beta _{i}}{\partial\tilde\lambda _{j}}\frac{
\partial\tilde\beta }{\partial\tilde\lambda _{i}}  \right) \frac{h^{5}}{v}
\nonumber \\ 
& & \!\!\!\!\! +\;  O\left(\frac{h^{6}}{v^2}\right) 
\eea
Here we use $\tilde{} $ to designate the SM quantities which generally differ from
the CW quantities.
Eq.(\ref{SMpot}) is  a ``low energy theorem'' for the SM
Higgs  potential in a limit in which the Higgs
boson is considered an approximate dilaton \cite{Grinstein}. 
We have retained a quintic term to remind the reader
that the standard model will have such terms, and beyond, 
owing to the RG running of $\lambda$.

Remarkably, we see that the Coleman-Weinberg potential
expanded about its minimum, $v$, depends {\em only upon $\beta$-functions and their 
derivatives at $v$}, \ie, is wholly determined by the renormalization group. Of course, we have
swapped $\lambda_1(v)$ for $v$, having used the extremal condition,
$\lambda_1(v) = -\beta_1(v)/4$,  to eliminate $\lambda_1(v)$.
The standard model has an input mass, and therefore we cannot eliminate the separate
$\lambda_1$ and $\beta_1$ dependences.
 We will use the improved potentials 
for comparson of the trilinear, quartic and qunitic terms  below.

\subsection{Role of the ``Improved Stress tensor''} 

Here we  emphasize the underlying canonical aspects of the dynamical
Coleman-Weinberg potential and renormalization group, in part to
give a formal basis to the idea of couplings that run with
field VEVs and a derivation of eq.(\ref{CWeqn}).  [A reader interested 
only in our phenomenological model
can skip this section and go directly to Section III.
The material is summarized here, and is
developed in greater detail in Appendix A.]

The  canonical stress tensor
of a real scalar theory with potential $V(\phi)$
is:
\bea
T_{\mu \nu } & = &  \partial _{\mu }\phi \partial _{\nu
}\phi -\eta _{\mu \nu }\left( \frac{1}{2}\partial _{\rho }\phi \partial
^{\rho }\phi -V(\phi) \right) 
\eea
 The  ``scale current,'' the Noether current associated
with scale symmetry, takes the form  $S_\mu = x^\nu {T}_{\mu\nu}$,
and divergence is given by  
$ \partial^\mu S_\mu = {T}_{\mu}^{\mu}$. These are defined
and derived in Appendix A.

The problem now arises that the canonical
stress tensor has a nonvanishing trace, $T_{\mu}^{ \mu }\neq 0  $, even
for a scale invariant theory. Yet, we see that the trace represents breaking of scale symmetry
since it is the divergence of the scale current. 
Therefore, an ``improved'' stress-tensor for scalar fields, $\widehat{T}_{\mu\nu}$ 
was introduced by Callan, Coleman and Jackiw \cite{CCJ}:
\bea
\widetilde{T}_{\mu \nu } 
& = &\frac{2}{3} \partial
_{\mu }\phi \partial_{\nu }\phi -\frac{1}{6}\eta _{\mu \nu }\partial _{\rho
}\phi \partial ^{\rho }\phi -\frac{1}{3}\phi \partial _{\mu }\partial
_{\upsilon }\phi 
\nonumber \\
& & 
%\!\!\!\!\!\!\!\!    \!\!\!\!\!\!\!\!  
+\frac{1}{3}\eta _{\mu \nu }\phi \partial ^{2}\phi +
\eta _{\mu \nu }V(\phi) 
\eea
The  scale current now  takes the form  $\hat{S}_\mu = x^\nu\widehat{T}_{\mu\nu}$.
and the trace  is found to be:  
\beq
\label{trace00}
\partial^\mu \hat{S}_\mu = \widehat{T}_{\mu}^{\mu} =4V(\phi)-\frac{d}{d\ln\phi} V(\phi)
\eeq
where we've used the equation of motion, $\partial^2\phi +V'(\phi)=0$.
Classically,  $\widehat{T}_{\mu\nu}$ is then traceless for a potential of
the form $V(\phi) \propto \lambda \phi^4$, where $\lambda$ is  a constant, 
reflecting the exact classical scale invariance 
of the theory. 
The improved stress-tensor is therefore required to discuss the scale symmetry
of scalar fields.  

The stress tensor is derived canonically from a scalar field action by performing
a coordinate variation, called a ``diffeomorphism.'' It can {\em alternatively} be
derived by performing
a variation of the background metric.
The improved stress tensor is generally viewed as emerging from a scalar field 
action, $S(\phi,g_{\mu\nu})$, which includes  the gravitational ``conformal
coupling term,'' $\half \xi\phi^2 R$ with $\xi=1/6$, and 
is given by $\widehat{T}_{\mu\nu}=-2 \delta S/\delta g_{\mu\nu}$.
However, there must exist, for symmetry reasons, another way of obtaining
the same improved stress tensor without considering the metric variation.

In Appendix A we provide a modified scalar field action  which 
 generates $\widehat{T}_{\mu\nu}$
while {\em maintaining a fixed flat-space metric}.  
Such a ``dual  derivation'' of
 $\widehat{T}_{\mu\nu}$  exists because of the
defining gauge symmetry of general relativity, 
general (Einstein) covariance:
if we simultaneously do the diffeomorphism and the {\em covariant} metric variation
(\ie, the particular metric variation under the diffeomorphism as dictated
by general covariance) then the
action must be invariant 
\footnote{ This dual derivation of the conserved current
is fundamental to any gauge theory,
and is analogous to the fact that the
electromagnetic current can be obtained by locally varying the vector potential
in the Dirac action, $\delta A_\mu$,
or by varying the phase of the electron wave-function, $\delta \psi = i\theta(x)\psi$.
Doing both at the same time with  $\delta A_\mu = \partial_\mu \theta$ is just
a gauge transformation, under which the Dirac action is invariant.}. 
The modifed action that generates
the improved stress tensor in flat space has an additional term, one that is
a total divergence, 
$\xi \partial^2\phi^2$.
This term, albeit non-topological, is similar to a $\theta$-term in QCD, undergoes a nontrivial variation 
when we perform the flat-space 
diffeomorphism. It generates a correction, $Q_{\mu\nu}$ 
which adds to the canonical stress tensor
and yields the improved stress tensor. 
The $\xi \partial^2\phi^2$ term remains a surface term
when the metric is non-flat, and it does not affect either the equations of motion, 
or any local variation of a non-flat metric.

When the matrix elements of the operator $\widehat{T}_{\mu\nu}$ are evaluated at
the quantum loop level for the classically scale invariant $\lambda \phi^4$ theory,
they are found to be nonzero at $O(\hbar)$, taking the operator value:
\beq
\label{tranom50}
 \widehat{T}_{\mu}^{\mu} =4V(\phi)-\frac{d}{d\ln\phi} V(\phi)
 = -\frac{\beta(\lambda)}{\lambda} V(\phi)
\eeq 
Eq.(\ref{tranom50}) is the {\em RG equation for the potential $V(\phi)$}.
The {\em rhs} is  the ``trace anomaly'' and it reflects the ${\cal{O}}(\hbar)$ breaking
of scale symmetry.  
[We carry out a Feynman 
loop evaluation of the trace anomaly in Appendix B. We also reproduce
the classic Coleman-Weinberg potential for massless scalar electrodynamics
 using the RG in Appendix C, and discuss some of its subtleties.]

Formally we  see that
we can represent the trace anomaly when the RG running
of $\lambda(\phi)$ as a function of $\phi$ is incorporated. 
By ``$\phi$'' we mean the VEV, or a  soft classical field
configuration.  We have from eq.(\ref{trace00})
when combined with  eq.(\ref{RG0}) with $\mu \rightarrow \phi$:
\beq
\label{tranom}
\widehat{T}_{\mu}^{\mu} =-\beta \phi^4 = -\frac{\beta(\lambda)}{\lambda} V(\phi)
\eeq
 The RG running of $\lambda$ with $\phi$
is essential to represent the anomalous result in the low
energy effective theory. This is much like the representation of the chiral
anomaly  by shifts in pNGB's, \eg, the pion or gauge fields, in a Wess-Zumino-Witten (WZW) term:
The running coupling constant scalar potential plays the analogous role
for scale symmetry anomalies that the  WZW term plays for chiral anomalies.
In the WZW term the axial anomaly is represented entirely
bosonically,  \ie, the
pion shift under a chiral transformation generates the axial anomaly. 
 We emphasize that the trace anomaly is an explicit, not spontaneous, breaking
of scale symmetry and there is no associated Nambu-Goldstone boson, \ie,
there need be no dilaton here \cite{Grinstein}.

\subsection{The trace anomaly is the ``anomalous dimension'' of the potential}

  Eq.(\ref{tranom50}) informs
us that  the ratio $\beta(\lambda )/\lambda $ is indeed the 
{\em anomalous dimension} of the potential.  This must become large,
 to manufacture mass from no mass. In fact, the condition that
the induced potential has an extremum, hence a local minimum, is precisely
that of eq.(\ref{oneten0}):
\beq
\label{anomdim}
\beta/\lambda = -4
\eeq
This result is true to all orders in a perturbation theory in $\hbar$ as we've seen Section II.B.
At the extremal point $\VEV{\phi} = v$  in field space the potential is converted to $D= 4-4=0$,
which corresponds to 
vacuum energy, \ie, a cosmological constant.

 If  dimensional transmutation is to occur, we see that 
the condition $\beta = -4\lambda$ implies
that an $O(\hbar)$ quantity, $\beta$, is being equated to
an $O(1)$ coupling constant $\lambda$.   This would seemingly violate
perturbation theory.  However, if there are 
additional coupling constants
beyond $\lambda$ that appear in $\beta$, \eg, $\beta \propto \hbar \lambda'{}^2$,
then the ratio $\lambda'{}^2/\lambda $ can easily be much greater than unity, while
maintaining perturbativity,
and the relationship eq.(\ref{anomdim}) can consistently occur.
This is at the heart of the Coleman-Weinberg phenomenon, as
emphasized in their paper \cite{Coleman:1973jx}.

\vskip 0.5in
\noindent
{\bf III. Phenomenological Model of the Higgs Boson Potential}
\vskip 0.2in

We now wish to apply the above apparatus to a model 
of the Higgs potential.
For simplicity, first consider the Higgs and top quark subset of the standard model:
\beq
{\cal{L}} = {\cal{L}}_{kinetic}+ g_{t}\bar{\psi}_Lt_RH + h.c. -\frac{\lambda}{2}\left( H^\dagger H\right)^2
\eeq
where $\psi = (t, b)$.
The one-loop RG equation for $\lambda$ is \cite{HLR}:
\beq
\label{model1}
\frac{d\lambda (\mu)}{d\ln (\mu)}=\beta(\lambda) = \frac{3}{4\pi^2}(\lambda ^{2}+\lambda
g_{t}^{2}-g_{t}^{4})
\eeq
where we neglect the electroweak couplings presently (we include these
below).

Let us  approximate $\beta$ as a constant in the SM.  
Note that, using the phenomenological values 
$ g_{t} \approx 1 $ and $ {\lambda} \approx {1}/{4}$,  we infer
from eq.(\ref{model1}):
\beq
\beta\approx -5.22\times 10^{-2}\qquad\makebox{ in the SM.}
\eeq
$\lambda$ is positive in the standard model, and $\beta$ is negative.
However, as we've seen in the previous section, if we want a CW effective potential
for the Higgs we require a negative
value of $\lambda$ and a positive $\beta$. 

Numerically, if a CW potential is to fit the
observed Higgs boson, we would require: 
\bea
\label{require}
\beta(v) & = & \frac{m_h^2}{v^2} = \frac{(126\GeV)^2 }{(174\GeV)^2}\approx 0.52\;\;\;\;
\makebox{and}\nonumber\\
  \lambda(v) & = & -\frac{\beta}{4} \approx -0.13.
\eea
  To make $\beta $ large and positive to $ O(\hbar)$ requires
more bosonic degrees of freedom \cite{bardeen2}.

Perhaps the simplest and most natural model for
a CW potential of the Higgs boson is to introduce a heavy second Higgs doublet,
$H_2$.  With a new Higgs doublet we have additional cross-coupling terms
$\sim (H_1^\dagger H_2)^2$. 
The most general classically scale invariant
potential with two massless Higgs doublets
and ``Weinberg-Glashow naturalness''
is well known \cite{WeinbergGlashow},\cite{CTH},\cite{Deshpande}:
\bea
\label{twodub}
V(H_1,H_2) & = & \frac{\lambda_1}{2}|H_1|^4 +  \frac{\lambda_2}{2}|H_2|^4 + {\lambda_3}|H_1|^2|H_2|^2 
\nonumber \\
& & \!\!\!\!\!\!\!\!\!\!    \!\!\!\!\!\!\!\!\!\! 
+ {\lambda_4}|H_1^\dagger H_2|^2+ \frac{\lambda_5}{2}\left[(H_1^\dagger H_2)^2 e^{i\theta} + h.c.\right]
\eea 
By judicious choice of parameters we can have one Higgs, $H_1$
develop a VEV, while $H_2$ remains dormant, \ie, no VEV.  

The potential of eq.(\ref{twodub}) has a ``Higgs parity'' symmetry
 $H_2\rightarrow -H_2$.  
Without couplings to fermions, additional Higgs doublets are
therefore stablized by this  symmetry,  and would become
stable dark matter.   Our present goal, however, is
to maintain reasonable visibility of the second doublet at the LHC
and we therefore require that $H_2$ can decay into visible final states.

Weinberg and Glashow  \cite{WeinbergGlashow} noted that such a parity
symmetry amongst the Higgs multiplets alone, can be broken
by couplings to fermions, but then
a larger reflection symmetry can exist where sets of the
coupled right-handed fermions
are also reflected, \eg, $\psi_R \rightarrow -\psi_R$, $H_2\rightarrow -H_2$.
The overall symmetry can be maintained if
we allow one new doublet per right-handed charge species in the
standard model.  This suppresses flavor changing neutral
Higgs boson couplings at tree-level  that would otherwise threaten such things as the 
small mass difference of the $K_LK_S$, but it now allows $H_2$ to decay
into the fermions it couples to. 
 
For  visibility at the LHC
the $H_2$ parity symmetry  must therefore be broken via coupling to fermions,
but the overall symmetry of Weinberg-Glashow maintained as much as possible.
However, $H_2$ is dormant, so the fermions coupled to it exclusively
cannot then get mass.  We therefore ultimately require some small breaking of
the overall Weinberg-Glashow symmetry. 

There are
two possibile schemes: {\bf (A)} We can have the $b_R$ couple, 
with possibly a large coupling constant, to $H_2$, respecting  Weinberg-Glashow symmetry,
but with its smaller SM coupling to $H_1$ allowing $m_b$ to be generated
by the $H_1$ VEV; {\bf (B)} we can add new ``centi-weak'' bosonic terms to the Higgs
potential that break the parity symmetry.

In scheme {\bf (A)} all quarks and leptons 
 couple  to  $H_1$ just as they do to the standard model
Higgs, and acquire mass via the $H_1$ VEV. 
We postulate that the $b$-quark, however, also has a large coupling $g_b'$ to $H_2$,
\beq
 g'_b\psi_L H_2^cb_R + h.c.
\eeq
where $\psi_L= (t,b)_L$ and $H^c=-\sigma_2 H^{*}$ (the choice of $b$-quark, as opposed
to other down quarks, is a modelling assumption, motivated to maintain a $(t,b)$ symmetry). 
We are therefore slightly violating the Weinberg-Glashow symmetry.
This then raises the question: ``Are we now in trouble with flavor constraints,
such as $b \rightarrow s+(g,\gamma)$?'' 
Not definitively, but the full analysis of the flavor physics of this scheme is beyond the scope
of the present paper.    There is, however, always an
escape route that was employed in ``topcolor'' models: we can
 assume flavor textures, such as 
in \cite{Buchalla}, where essentially the CKM matrix arises via the ``up''
type quarks, and the Higgs couplings of ``down'' types are diagonal.
This suppresses any large flavor changing neutral
Higgs mediated transitions. In any case, a more detailed 
analysis of flavor constraints is warranted.  Certainly
the model survives in the $g_b'\rightarrow 0$ limit where
the Weinberg-Glashow symmetry is recovered, but  gluon fusion
associated production of $H_2$ at the $\sim 100$ fb level will then turn off,
while EW production at the $\sim 1$ fb level remains (see IV.(A)).  

Alternatively, in scheme {\bf (B)} all $+2/3$ quarks and leptons 
 couple  to  $H_1$ as in the standard model, and acquire mass via the $H_1$ VEV,
but we have
no coupling of $-1/3$ quarks to $H_1$ to the $b$-quark. Here the  down
quarks coupled only to $H_2$, which maintains the Weinberg-Glashow symmetry.
We then break this symmetry by introducing a 
bosonic interaction:
\beq
\label{newint}
\frac{\lambda^{_{^{\prime }}}}{2}\left( H_1^{\dagger }H_1 \right)\left( H_1^{\dagger
}H_{2}\right) +h.c.
\eeq
(of course, this new interaction would be induced
by fermion loops involving down-quarks, if they coupled to both
$H_1$ and $H_2$).  
Since this interaction also breaks Weinberg-Glashow naturalness, we therefore
expect $\lambda'$ to be small.   

The bosonic interaction of eq.(\ref{newint})
leads to an interesting effect that may explain 
the flavor hierarchy between $+2/3$ and $-1/3$ charge species.  
When the Higgs $H_1$ acquires a VEV,
$\VEV{H_1} = (v, 0)$
it induces a tadpole interaction to the neutral component
of $H_2$,
\beq
\frac{\lambda^{\prime }}{2\sqrt{2}} v^{3}H^0
\eeq
where $H_2= ((H^0+iA^0)/\sqrt{2}, H^-)$.   $H_2$ is initially
dormant and will acquire a large
positive mass from the $H_1$ VEV,  $\sim M^2H_2^\dagger H_2 $.
But, through the tadpole, we obtain a small induced VEV for $H^0$:
\beq
\VEV{H^0} = \frac{\lambda ^{\prime }}{\sqrt{2}M^2 }v^{3}
\eeq
The down quarks will then have small induced masses, $\sim \lambda' v^3/M^2 $
and $\lambda' \sim  O(10^{-2})$.
The interaction eq.(\ref{newint}) also splits the neutral and charged
members of the dormant doublet.  

In scheme {\bf (B)} the
dormant Higgs will have decay modes via eq.(\ref{newint}) such as $H^0\rightarrow 3 h$,
and/or $H^0\rightarrow 2 h + (h^*\rightarrow b\bar{b})$, \etc, 
and radiative modes $H^\pm \rightarrow W^\pm + 2h +h^*$, \etc.  These
are interesting modes to search for, but their detailed analysis is beyond
the scope of the present paper and require further study.  
We will focus here upon the phenomenology
of scheme {\bf (A)}.

The general RG equations for two-doublet models are given
in ref.\cite{HLR}.  We introduce fermionic couplings and
we choose as a starting point
 Model IV as defined in \cite{HLR}. We assume operationally 
that $H_1$ couples to the top quark
 via $g_t$ 
and $H_2$ couples to the $b$ quark via $g'_b$ (we ignore all
other smaller Higgs-Yukawa couplings),
\beq
g_t\psi_L H_1 t_R + g'_b\psi_L H_2^cb_R + h.c.
\eeq
where $\psi_L= (t,b)_L$ and $H^c=-\sigma_2 H^{*}$.

With the additional $\lambda_i$ of eq.(\ref{twodub}) the RG equations
become \cite{HLR}:  
\bea
\label{RGs}
16\pi^2 \frac{d\lambda_1 (\mu)}{d\ln (\mu)}
& = & 12\lambda_1^{2}+4\lambda_3^2+4\lambda_3\lambda_4+2\lambda_4^2+2\lambda_5^2 \nonumber \\
& & -3\lambda_1 (3g_2^2+g_1^2) +\frac{3}{2}g_2^4 + \frac{3}{4}(g_1^2 + g_2^2)^2 \nonumber \\
& & + 12\lambda_1 g_t^2-12 g_t^4
\\
16\pi^2 \frac{d\lambda_2 (\mu)}{d\ln (\mu)}
& = & 12\lambda_2^{2}+4\lambda_3^2+4\lambda_3\lambda_4+2\lambda_4^2+2\lambda_5^2 \nonumber \\
& & -3\lambda_2 (3g_2^2+g_1^2) +\frac{3}{2}g_2^4 + \frac{3}{4}(g_1^2 + g_2^2)^2 \nonumber \\
& & + 12\lambda_2 g'_b{}^2-12 g'_b{}^4
\\
16\pi^2 \frac{d\lambda_3 (\mu)}{d\ln (\mu)}
& = & (\lambda_1 +\lambda_2)(6\lambda_3+ 2\lambda_4)+
4\lambda_3^2 +  2\lambda_4^2 +2\lambda_5^2 \nonumber \\
& & -3\lambda_3 (3g_2^2+g_1^2) +\frac{9}{4}g_2^4 + \frac{3}{4}g_1^4   
- \frac{3}{2}g_1^2 g_2^2 \nonumber \\
& & + 6\lambda_3(g_t^2+ g'_b{}^2)-12 g_t^2 g'_b{}^2
\\
16\pi^2 \frac{d\lambda_4 (\mu)}{d\ln (\mu)}
& = & 2(\lambda_1 +\lambda_2)\lambda_4+
4(2\lambda_3 + \lambda_4)\lambda_4 +8\lambda_5^2 \nonumber \\
& & -3\lambda_4 (3g_2^2+g_1^2) +3g_1^2 g_2^2 -12 g_t^2 g'_b{}^2
\\
16\pi^2 \frac{d\lambda_5 (\mu)}{d\ln (\mu)}
& = & \lambda_5[ 2(\lambda_1 +\lambda_2)+
8\lambda_3 + 12\lambda_4 \nonumber \\
& & -3(3g_2^2+g_1^2)  + 2(g_t^2+ g'_b{}^2)]
\eea

We've analyzed many variations
of this model with $\lambda_3$, $\lambda_4$, and $\lambda_5$ 
all active.  Presently we'll discuss only
 the  simplest case
with $\lambda_5=\lambda_4=0$.   $\lambda_5$ 
breaks a global symmetry, $H_1\rightarrow e^{i\theta}H_1$, 
$H_2\rightarrow e^{-i\theta}H_2$, 
and is therefore multiplicatively renormalized. Hence,
it remains zero once set to zero, and this is evident
in the RG equation above for $\lambda_5$. 
Moreover, in the absence of $\lambda_5$ and ignoring the
the  $SU(2)\times U(1)$
gauge fields, we
see that $\lambda_4$ breaks a larger symmetry,
$SU(2)\times SU(2) \rightarrow SU(2)$, and it too is then multiplicatively
renormalized. If $\lambda_4$ is
set to zero at some high scale, it therefore remains reasonably small  and
can be ignored.

Let's estimate the required effect of $\lambda_3$
needed to create the Coleman-Weinberg potential for $H_1$. 
We have at one-loop order from eq.(\ref{CWfull}):
\beq
m_h^2= v^{2}  \beta_1 \qquad \makebox{and,} \qquad  \lambda_1 =-\frac{1}{4}\beta_1
\eeq
hence:
\beq
\beta_1=\frac{m_h^2}{v^2} \approx 0.524 \qquad  \lambda_1 =-0.131
\eeq
From eq.(\ref{RGs}) we also have:
\bea
16\pi^2 \beta_1
& = & 12\lambda_1^{2}+4\lambda_3^2+4\lambda_3\lambda_4+2\lambda_4^2+2\lambda_5^2 + 12\lambda_1 g_t^2
\nonumber \\
& &\!\!\!\!\! \!\!\!\!\! \!\!\!\!\! \!\!\!\!\! \!\!\!\!\!
 -3\lambda_1 (3g_2^2+g_1^2) +\frac{3}{2}g_2^4 + \frac{3}{4}(g_1^2 + g_2^2)^2 
%\nonumber \\
%& & 
-12 g_t^4 
\nonumber\\
&  & \!\!\!\!\! \!\!\!\!\! \!\!\!\!\! \!\!\!\!\! \!\!\!\!\! \approx 0.0253\lambda_3^2 - 0.0668
\eea
which yields:
\beq
\lambda_3 \approx  4.83
\eeq
(We use $m_h = 126$ GeV, $v=174$ GeV, $m_t = 173.5$ GeV, so $g_t=0.997$;
also $g_2^2=0.425$, $g_1^2=0.127$).
While this is a rather large coupling, it is still perturbative, as its
contribution to the $\beta_i \lta 1 $.

When the Higgs, $H_1$, acquires its VEV the $\lambda_3|H_1|^2|H_2|^2 $ term of the potential, eq.(\ref{twodub}),
will induce a mass for $H_2$,  $M_{H_2}^2 =\lambda_3v^2 $.
We require that the dormant doublet $H_2$ have a positive $M_{H_2}^2$
and therefore,  $\lambda_3$
is positive.  We thus estimate:
\beq
 M_{H_2}\approx \sqrt{4.83}\times
(174)\;\makebox{ GeV}\;\; \approx 382 \makebox{ GeV}
\eeq

With such a large $\lambda_3$ we can improve the prediction by including the two-loop
effect of  eq.(\ref{CWfull}). 
The Higgs mass is given by
\beq
m_h^2= v^{2} \left( \beta_1  + \frac{1}{4}\beta_3\frac{\partial \beta_1}{\partial\lambda_3} \right)
\eeq
where the second term arises at two-loop level.
We can use the leading dependence upon the large $\lambda_3$ in the
last term. From eq.(\ref{RGs})
\beq
\beta_3 \approx \frac{\lambda_3^2}{4\pi^2} \qquad 
\frac{\partial \beta_1}{\partial\lambda_3}
\approx \frac{\lambda_3}{2\pi^2}
\eeq
hence,
\beq
0.524 = \left( \beta_1  + \frac{\lambda_3^3}{32\pi^4} \right)
\eeq
(note that ${\lambda_3^3}/{8\pi^4}\approx 0.0362$ which is the scale
of these higher order corrections is small).
Solving again for $\lambda_3$, we now obtain:
\beq
\lambda_3 \approx  4.68\qquad M_{H_2}\approx \sqrt{4.68}\times 174\;\makebox{ GeV} = 376 \;\makebox{ GeV}.
\eeq

Of course, the large $\lambda_3$ leads to a ``UV 
challenge'' for this scheme.  Since $\lambda_3$ is large,  it's RG running
into the UV  leads to a Landau pole.
Indeed, we see this from a  numerical integration of eqs.(\ref{RGs})
in Figure (4).
We have considered the effects of the additional couplings, $\lambda_4$ and
$\lambda_5$ and have not found an elegant or simple remedy to this problem
without a significant extension of the model.  

We note that we can somewhat improve the UV behavior 
of this scheme by considering
$H_2$ to be a QCD color triplet, $(3,2,Y=1/3)$, and $Q=I_3+Y/2$.  In this structure
then $(H_1, H_2)$ form a bosonic generation, similar to a lepton-quark
generation, with $H_2=(H^{+2/3}, H^{-1/3})$.  
$H_2$ can then couple to a quark-lepton
combination, e.g. $g_{\nu q}\bar{\psi}H_2\nu_R$
or $g_{\ell q}\bar{\psi}H^c_2\ell_R$. $H_2$ thus
becomes a ``lepto-quark.''   The $\nu_R$ case is
intriguing, as we would integrate it out as in neutrino Majorana
masses, and  $H_2$ then becomes dark matter.

We can drastically modify the scheme to push the Landau pole upwards in energy scale,  
by imbedding $SU(2)\times U(1) \rightarrow SU(2)\times SU(2)\times U(1)$
 at some high energy scale, $\Lambda$, below the
Landau pole.  This is analogous to ``top-flavor'' models \cite{topflavor},
and can be done in a flavor democratic way. The
effect is to replace $g_2$ with a larger $g_2'= g_2/\sin(\chi)$ in the RG equations.
This improves the  UV behavior.  Landau poles generally reflect compositeness of fields \cite{BHL}. 
The compositeness
conditions are associated with the vanishing of wave-function normalization
constants, $Z_H$.

%\begin{figure}[h]
%\includegraphics[bb = 0 0 100 100 ]{UV1.pdf}
%\caption{UV running of the Dormant Higgs model}
%\end{figure}
\begin{figure}[t]
\vspace{6cm}
\includegraphics{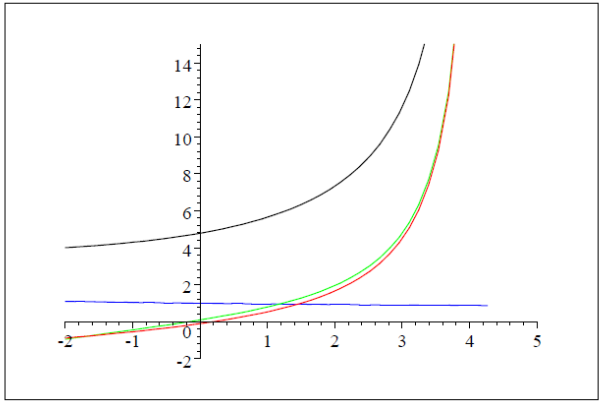}
\vspace{-0cm}
\caption[]{ UV running of the dormant Higgs model, $\lambda_i$  vs. 
$\ln ( \mu/ v_{weak} )$ (black-$\lambda_3$, red-$\lambda_1$, green-$\lambda_2$,
blue-$g_{top}$). This shows Landau singularity at $\ln(\mu/v_{weak})
\sim 3\sim 4$, where $v_{weak}=175$ GeV, or $\mu \sim 3\sim 10$ TeV.}
\label{xroots}
\end{figure}

%\newpage
\vskip 0.2in
\noindent
{\bf IV. Phenomenology of $H_2$ }

\vskip .1in
\noindent
{\bf (A) Production and Decay of the Dormant Higgs} 
\vskip .1in 
  
We have carried out estimates of decay widths and production cross-sections 
of the dormant doublet, $H_2$, using CalcHEP. We've adapted the ``inert doublet model,'' with inclusion of the Yukawa couplings
to the $b$-quark (see \cite{Semenov}).
Since our goal was to maintain ``maximal visibility'' of the new bosons that allow a Coleman-Weinberg potential for the Higgs, $H_2$ is necessarily coupled to the $SU(2)\times U(1)$ gauge fields of the
standard model.  The doublet does not have a VEV, but 
(ignoring fermion couplings ) the neutral components, which we denote as $H^0$ and $A^0$,
are pair-produced via $\gamma^*$, $Z^*$; the charged components $H^\pm$  are likewise
pair-produced via $W^\pm$.  

We follow the conventional nomenclature of the SUSY two doublet schemes, but we emphasize that the minimal
theoretical scheme maintains an approximate degeneracy between $H^0$ and $A^0$, and $H^\pm$,
hence the doublet is defined as $H_2=((H^0 + iA^0)/\sqrt{2}, H^-)$. 
The degeneracy is broken by the Yukawa couplings to matter, $g_b{}'(\bar{t}, \bar{b})^T_L H^C_2 b_R + h.c.$, where $H^C = i\sigma_2 H^*$. New quartic interactions that lift
the degeneracy by way of the normal Higgs boson VEV (which is the neutral member of $H_1$)
will be induced by fermion loops.
Throughout we have assumed the degenerate doublet with $M_{A^0}= M_{H^0} = M_{H^\pm} = 380 $ GeV$/c^2$.

\begin{figure}[t]
\label{f1}
\vspace{6.0cm}
\includegraphics{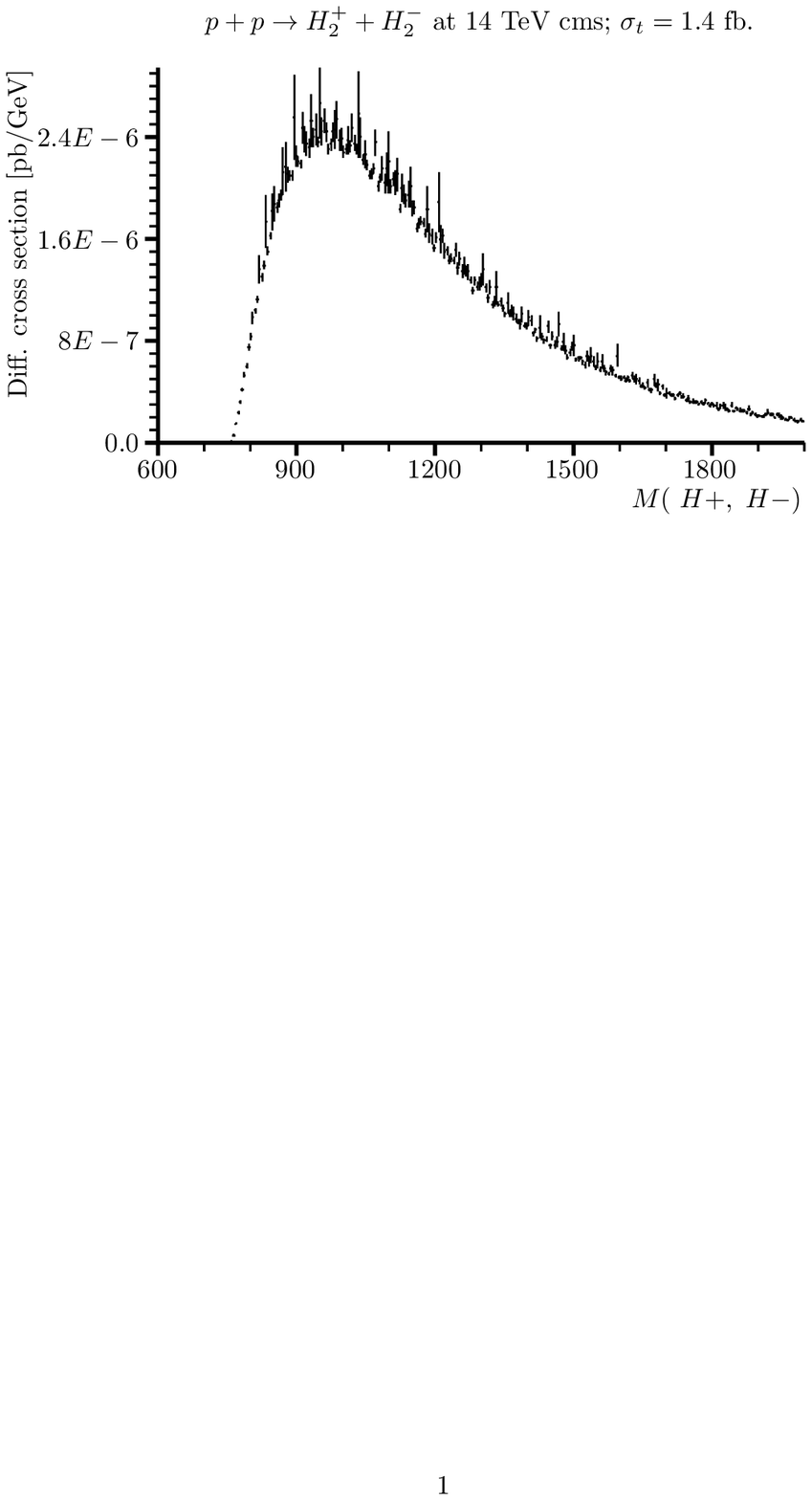}
\vspace{-0.5cm}
\caption[]{ $H^+H^-$ production at LHC.}
\label{xroots}
\end{figure}

\begin{figure}[t]
\label{f2}
\vspace{6.0cm}
\includegraphics{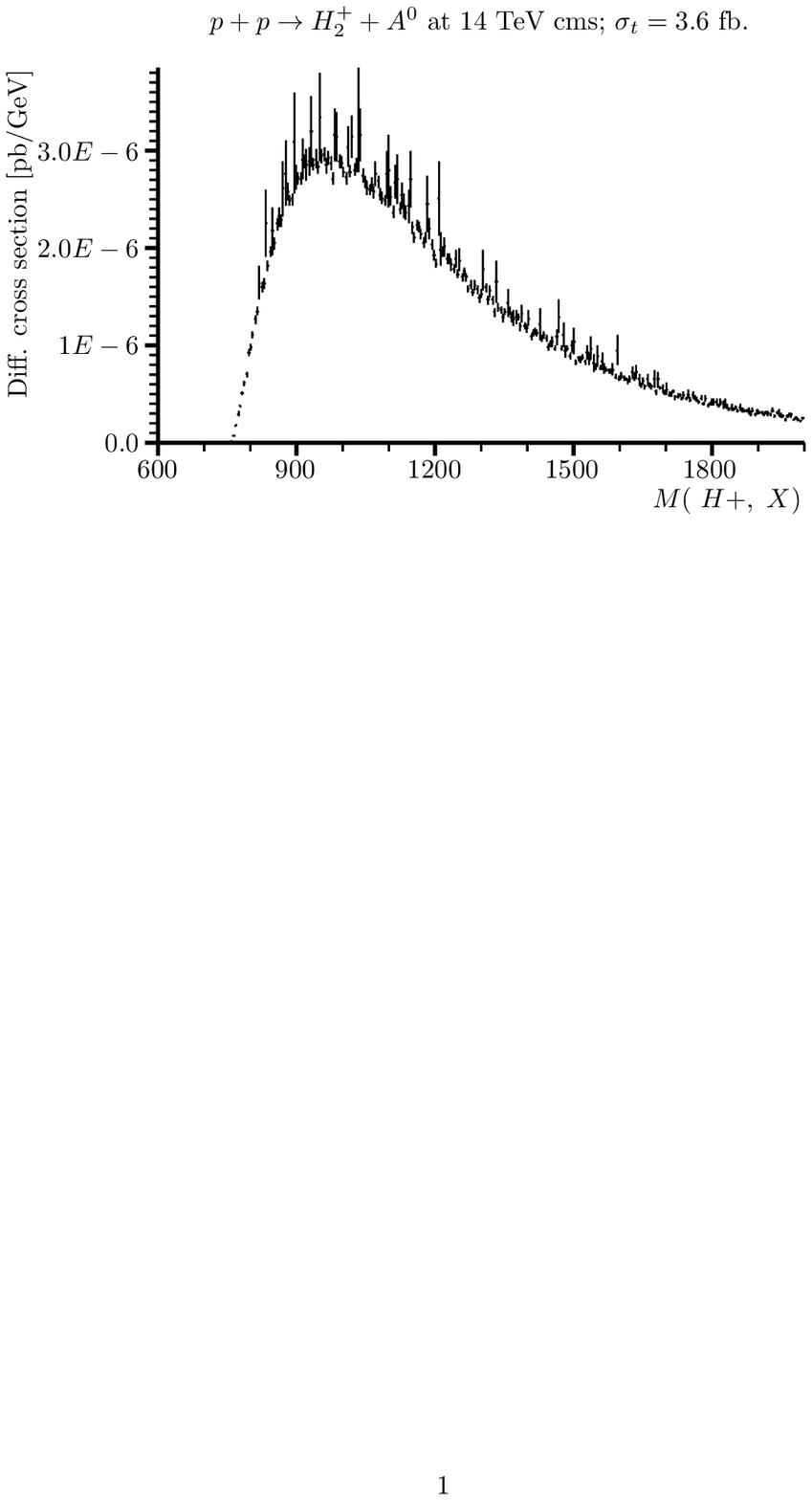}
\vspace{-0.5cm}
\caption[]{ $H^+A^0$ production at LHC.}
\label{xroots}
\end{figure}

%%%%%%%%%%%%%%%%%%%%%%%%%%%%%%%%%%%%%%%%%%%%%%%%%%%%%%%%%%%%%%%%%%%%%%
\begin{table*}
\label{widths}
\begin{center}
\caption{Predicted decay widths and production cross-sections for the 
dormant Higgs bosons. We used CalcHep, and  production runs
CTEQ61 proton structure functions, $1.64\times 10^5$ calls. All cross-sections are
evaluated at $14$ TeV cms energy with the mass of $H_2$ doublet
set to $380$ GeV$/c^2$.  Model dependent processes have rates
or cross-sections that are indicated as $\propto (g_b')^2$.
}
\begin{tabular}{llll} 
\hline
\hline
Process & &  value  &  comments \\
\hline
\hline
 $\Gamma(H^+\rightarrow t+\bar{b}) = \Gamma(H^-\rightarrow b+\bar{t}) $ &  & $14.5\; (g_b')^2 \pm 5\times 10^{-5} \% $ GeV    \\
\hline
$\Gamma(H^0 \rightarrow {b}+\bar{b}) = \Gamma(A^0\rightarrow {b}+\bar{b}) $ &  & $5.67\;  (g_b')^2 \pm 5\times 10^{-5} \% $ GeV     \\
\hline
$\Gamma(H^0\rightarrow 2h, 3h)= \Gamma(A^0\rightarrow 2h, 3h) $ &    & & absent in model \\
\hline
\hline
 $ pp \rightarrow (\gamma, Z^0)\rightarrow H^+ H^- $ &  &  $\sigma_t =  1.4\;$ fb    \\
\hline
 $ pp \rightarrow (\gamma, Z)\rightarrow H^0 H^0 $ &  & & absent in model \\
\hline
  $ pp \rightarrow (\gamma, Z)\rightarrow A^0 H^0 $ & & $\sigma_t =  1.3 $  fb \\
\hline
 $ pp \rightarrow (\gamma, Z)\rightarrow A^0 A^0 $ & & & absent in model \\
\hline
 $ pp(gg) \rightarrow h\rightarrow H^0 H^0 $ or $A^0A^0$ &  & $ \sigma_t =  1.7\times 10^{-5} $  fb   \\
\hline
 $ pp \rightarrow W^+ \rightarrow H^0 H^+ $ & & $\sigma_t = 1.8 $   fb \\
\hline
 $ pp \rightarrow W^+ \rightarrow A^0 H^+ $ & & $\sigma_t = 1.8 $   fb \\
\hline
 $ pp \rightarrow W^- \rightarrow H^0 H^- $ &  &  $\sigma_t =  0.74 $  fb   \\
\hline
 $ pp \rightarrow W^- \rightarrow A^0 H^- $ &  &  $\sigma_t =  0.74 $  fb   \\
\hline
\hline
 $ pp \rightarrow b+\bar{b}+ H^0 $ or $A^0$   &  & $\sigma_t =  1.8 \; (g_b')^2$ pb $\pm 2.4\% $ &   No 
$p_T$ cuts \\
\hline
 & &  $\sigma_t = 67\; (g_b')^2 $ fb $ \pm 5\%$  &   $  p_T(b)$ and $ p_T(\bar{b})>50 $ GeV \\
\hline
  &   & $ \sigma_t = 9.6 \; (g_b')^2 $ fb $\pm 3.5\%$ &   $  p_T(b)$ and $ p_T(\bar{b})>100 $ GeV \\
\hline
\hline
 $ pp \rightarrow  t+\bar{b}+ (H^-) $ & &  $\sigma_t =  220 \; (g_b')^2$ fb &   No cuts\\
\hline
  &   & $\sigma_t =  44\;  (g_b')^2 $ fb  &   $p_T(t) $, $p_T(\bar{b})$ $>$ $50$ GeV\\
\hline
  &   & $\sigma_t =  14\; (g_b')^2 $ fb   &   $p_T(t) $, $p_T(\bar{b})$ $>$ $100$ GeV\\
\hline
 $ pp \rightarrow  \bar{t}+{b}+ (H^+) $ &  &  $\sigma_t =  270\; (g_b')^2 $ fb  &   No cuts\\
\hline
  &   & $\sigma_t =  46\; (g_b')^2 $ fb   $p_T(\bar{t}) $ &  $p_T({b})$ $>$ $50$ GeV\\
\hline
  &   & $\sigma_t =  14\; (g_b')^2 $ fb  $p_T(\bar{t}) $ &  $p_T({b})$ $>$ $100$ GeV\\
\hline
\hline
\end{tabular}
\end{center}
\label{tbl:width_exp}
\end{table*}

In principle $H_2$ could exist with no coupling to matter.  However, as described in the previous
section, we will allow a large $O(1)$ coupling $g_b'$ to the $b$-quark.  As such, the neutral 
$H^0$ and $A^0$ can then be pair produced in association with $b\bar{b}$, and the charged
$H^\pm$ in association with $b\bar{t}$ and $t\bar{b}$. 

The decay widths $\Gamma(H^+\rightarrow t+\bar{b}) = \Gamma(H^-\rightarrow b+\bar{t}) $,
and $\Gamma(H^0 \rightarrow {b}+\bar{b}) = \Gamma(A^0\rightarrow {b}+\bar{b}) $ are then generated and  computed in Table I. Note that in our minimal scheme a parity $H_2\rightarrow -H_2$
that would make the $H_2$ components stable is broken only by the Yukawa coupling to $(t,b)_L\bar{b}_R$.
Therefore, at tree level the decays $\Gamma(H^0\rightarrow 2h, 3h)$, $ \Gamma(A^0\rightarrow 2h, 3h) $
are absent in the model.  The decay widths are of order $\sim 10$ GeV for $g_B'\sim O(1)$.
The distributions for these processes are indicated in Figs.(5,6).

The SM gauge production cross-sections are computed for the LHC RUN-II at
$\sqrt{s} = 14$ TeV. While small, $\sim O(1)$ fb, these may be observable with $\sim 100$ fb$^{-1}$ 
of data and judicious cuts. 

We have also computed the model dependent 
($\propto g_b'{}^2$ associated production rates for $ pp \rightarrow  b+\bar{b}+ (H^0, \;A^0) $, $ pp \rightarrow  t+\bar{b}+ (H^-) $  and $ pp \rightarrow  \bar{t}+{b}+ (H^+) $.  These are predominantly gluon fusion processes at the LHC.  We have applied various $p_T$ cuts as indicated
on final state particles, but we have not done a detailed signal/background analysis,
requiring more careful detector dependent study.

\vskip .5in  
\noindent  
{\bf (B) Trilinear, Quadrilinear, and Quintic  Higgs Coupling}
\vskip .2in  

A characteristic feature of the Coleman-Weinberg potential
 is that the trilinear, quadrilinear and quintic 
Higgs couplings differ dramatically from
that of the standard model \cite{CWCTH}, \cite{Dermisek}.  

With the standard model polynomial potential we have:
\bea
V_{SM}(H) 
& = &  \hat\lambda v^2h^2 +\frac{\hat\lambda}{\sqrt{2}} v h^3+\frac{1}{8}\hat\lambda h^4
+ \frac{1}{24\sqrt{2}v}\hat{\beta} h^5 
\nonumber \\
&=&  
\half m_h^2 h^2  +\frac{m_h^2}{2\sqrt{2}v} h^3+\frac{m_h^2}{16v^2} h^4
+ \frac{1}{24\sqrt{2}v}\hat{\beta} h^5
\nonumber \\
& & + \; ...
\eea
From the Coleman-Weinberg potential expanded to quintic order, keeping
the leading two-loop $\lambda_3$ terms, we have:
\bea
V_{CW}(H)  & = & \half m_h^2 h^2  
%\nonumber \\
%& & 
+\frac{5}{6\sqrt{2}v} h^3\left(\beta_1 + \frac{9}{20}\beta_3\frac{\partial \beta_1}{\partial\lambda_3}\right)
\nonumber \\
& & 
%\!\!\!\!\! \!\!\!\!\! \!\!\!\!\! \!\!\!\!\! \!\!\!\!\! 
+\frac{11}{48v^2} h^4\left( \beta_1 + \frac{35}{44}\beta_3\frac{\partial \beta_1}{\partial\lambda_3} \right)
\nonumber \\
& & 
+\frac{1}{40\sqrt{2} v} h^5\left( \beta_1 + \frac{25}{12}\beta_3\frac{\partial \beta_1}{\partial\lambda_3}\right)+ ...
\eea
The ratios of the Coleman-Weinberg to standard model trilinear, quadrilinear and quintic terms are then:
\bea
\makebox{trilinear} & = &  \frac{5}{3}\left(1 + \frac{v^2}{5m_h^2}\frac{\lambda_3^3}{8\pi^4}\right)
\; \approx 1.75
\nonumber \\
\makebox{quadrilinear} & = &  \frac{11}{3}\left(1 + \frac{35v^2}{44m_h^2}\frac{\lambda_3^3}{8\pi^4}\right)
\; \approx 4.43
\nonumber \\
\makebox{quintic} & = & 
 \frac{3}{5}\left(\frac{\beta_1}{\hat\beta} + \frac{25}{12\hat\beta}\frac{\lambda_3^3}{6\pi^4}\right)\; \approx -8.87
\nonumber \\
\eea
where $\hat\beta\approx -0.0522$ is the SM $\beta$-function for $\lambda$.
The leading terms in the above,  independent of the new bosonic physics $\sim \lambda^3_3$  
are valid to  $O(\hbar)$, while the  $\sim \lambda^3_3$ are the leading largest $O(\hbar^2)$
terms.

The sensitivity at the LHC Run II is expected
to be comparable to these departures from the standard model,
and in future high-luminosity mode these effects should be observable with precision.
Future $e^+e^-$ Higgs factories would have sensitivity at the level
of $\sim 10\%$ or better.

While this is a model independent check on the ``Higgs with CW potential''
scheme we are considering, it is not the case in other models.  For example, in
\cite{Hambye1} the second doublet is {\em inert}, and does not couple to
the standard model $SU(2)\times U(1)$. The second doublet $H_2$ couples
to a new $SU(2)$ gauge interaction and develops a large VEV. The new $SU(2)'$
gauge fields together with $H_2$  become a dark matter ecosystem.  In this
model the Higgs acquires a negative mass$^2$ via a negative $\lambda_3$ (``Higgs portal
interaction''), and the
resulting $H_1$ potential is classical.  There would be no large trilinear effect
in this model, and it is presumably hard to test this scheme at the LHC.

\vskip .5in
\noindent
{\bf V.  Fermionic Origin of a CW Higgs Potential }

\vskip .2in

Remarkably, the full structure of eq.(\ref{CWfull}) admits an
alternative origin of a Coleman-Weinberg potential for the Higgs boson {\em via fermions}. 
This exploits the two-loop contribution to the $h^2$ term. 
We will presently give a schematic discussion of this possibility,
but it requires  more model building effort which we will
pursue elsewhere \cite{Ross}.

Suppose there exists a new fermion $SU(2)_L$ doublet
$\psi_L=(T, B)_L$, and a pair of singlets $(T_R, B_R)$.  Hence
the $\psi_L$ fermion couples to the standard model $SU(2)_L\times U(1)$,
gauge bosons while $\psi_R$  has only  $U(1)$ weak hypercharges. We further assume these
new fermions are ``hyperquarks,'' forming an ${\bf [N_c]}$ fundamental
representation, coupled to an unbroken strong gauge interaction,
$SU(N_c)$, ``hypercolor,''  with coupling constant $\tilde{g}$.
We'll ignore the issue of anomaly cancellation presently.

We assume the $U(1)$ charges are so chosen that the interaction
with a massless Higgs boson can occur as:
\beq
\label{HY}
g\bar{\psi}_L H T_R + g\bar{\psi}_L H^c B_R
\eeq
and we'll assume a common Yukawa coupling (we'll work
in the approximation of custodial $SU(2)$ invariance).

The Higgs boson is massless but has the usual quartic
potential with RG equation for $\lambda$
dominated by the Higgs-Yukawa $g^4$ term \cite{HLR}:
\bea
\label{grun}
\frac{d\lambda }{d\ln(\mu)} &  = & \beta_1 = \frac{1}{4\pi^2}\left( 3 \lambda^2 + 2N_c \lambda g^2 -  2N_c g^4\right)
\nonumber \\
& & \approx  -\frac{N_c}{2\pi^2}  g^4
\eea
where we've neglected the top Yukawa, and electroweak contributions.
Likewise, the RG equation for the Yukawa coupling $g$ takes the form \cite{HLR}:
\beq
\frac{d g }{d\ln(\mu)} = \beta_g = \frac{g}{16\pi^2} \left(2N_cg^2- (N_c^2-1)\tilde{g}^2\right)
\eeq

The Higgs potential can develop a dynamical minimum for a VEV, $v$, provided
that:
\beq
\label{anomdim100}
\frac{\beta_1(v)}{\lambda(v)} = -4
\eeq
Previously we have studied that case where $\beta_1(v)>0$ and $\lambda(v)<0$.
We will now discuss a case with the new fermions in which $\beta_1(v)< 0$ and $\lambda(v)> 0$,
and we assume at some $v$ that eq.(\ref{anomdim100}) holds.

A stable  minimum of the potential requires $m_h^2 > 0$. From
eq.(\ref{CWfull}), including the two-loop term, we see that:
\beq
m_h^2 = v^2\left(\beta_1 + \frac{1}{4}\beta_g\frac{\partial \beta_1}{\partial g} \right) >0
\eeq
Using the approximate form of eq.(\ref{grun}) we have:
\beq
 \frac{1}{4}\frac{\partial \beta_1}{\partial g} =-\frac{N_c}{2\pi^2}  g^3
\eeq
hence,
\bea
m_h^2 & \approx & v^2 \left(-\frac{N_c}{2\pi^2}  g^4 \right)
\left(1 + \frac{1}{16\pi^2} \left(2N_c g^2- (N_c^2-1)\tilde{g}^2\right)  \right)
\nonumber \\
& \approx & v^2\beta_1\left(1 + \frac{\beta_g}{g} \right) 
\eea
The condition that $m_h^2$ is positive is now the simultaneous conditions of
eq.(\ref{anomdim100}) and:
\beq
\label{anomdim101}
 \frac{\beta_g}{g} < -1
\eeq
The latter condition states that the anomalous dimension of the Higgs-Yukawa interactions
eq.(\ref{HY}) is less than $-1$ and thus the dimensionality
of this operator is reduced to $D < 4-1 = 3$.
Eq.(\ref{anomdim101}) can be realized by
\beq
 \frac{\tilde{\alpha}}{4\pi} > \frac{1}{(N_c^2-1)} + \frac{N_c g^2}{8\pi^2(N_c^2-1)} 
\eeq
where $\alpha_g = \tilde{g}^2/4\pi $.  The subsequent running of the hypercolor
$\alpha_g$ into the infrared is model dependent. With $N_f$ additional 
inert fermion flavors (not coupled to $SU(2)\times U(1)$),  $\alpha_g$ will blow up
at a scale $\Lambda_{HC} \sim v\exp(-6\pi/(11N_c-2N_f)\hat{\alpha}(v))$, and confine.
 This could in principle be a
walking theory.  With the minimal $(T, B)$, $(N_f=0)$ we see that $\Lambda_{HC} \sim 0.6 v $.
We prefer a limit $\lambda_{HC} << v$ so that the masses of the $(T,B)$ states are far
above the confining scale of hypercolor, and no chiral condensates are formed.

The Higgs mass at the minimum is therefore given by:
\bea
m_h^2 & = & v^2 \frac{N_c g^4}{2\pi^2}  \left(\left|\frac{\beta_g}{g}\right|-1 \right) 
\eea
If we assume $N_c = 3\; (4)$ and $\left|\frac{\beta_g}{g}\right|-1 = \kappa \approx 1$
we find that the masses of the new hyperquarks are $M \approx 236/(\kappa)^{1/4} \;( 219/(\kappa)^{1/4} )$ GeV.

Note that these objects would appear effectively as new leptons since they do not
interact with ordinary QCD $SU(3)_c$, and are not produced in gluon fusion.
Hypercolor could be QCD-like and confine at some scale $\lambda_{HC}$. We assumed that this is less that the inferred
Higgs-induced masses, $\sim 230$ GeV of $(T,B)$; therefore the resulting states are analogues of heavy quark-onium
boundstates in QCD, and there are no light pNGB's. 

 The new states will be pair produced via a single $Z^*$ or $W^*$
at a threshold $\sim 2M$, into a single $\bar{Q}Q$
heavy meson, (plus recoil jets of conventional quarks). 
The heavy $\bar{Q}Q$ decays into electroweak gauge bosons.
Open $\bar{Q}+ Q$ requires the the recombination into
pairs of mesons,  $\bar{Q}Q + \bar{Q}Q $ and a threshold
energy of $4M$.

Using fermions to engineer the CW potential
may allow a much more natural UV completion than the
bosonic $H_2$ model presented above.
The detailed model structures,
production and decay phenomenology is beyond the scope of the
present discussion.  Our interest here is to give a proof of principle of
the phenomenon of fermion-driven Coleman-Weinberg potentials.

%\newpage

\vskip 0.5in
\noindent
{\bf VI. Conclusions}
\vskip .2in

We have discussed the possibility that the electroweak scale is
a quantum phenomenon, \ie, that it arises via particle loops leading to
a perturbative Coleman-Weinberg potential. We have developed the renormalization
group formalism for the Coleman-Weinberg potential, and its
relationship to the trace anomaly of the improved stress tensor
for scalar fields.  An expansion of the CW potential about
its minimum, valid to all orders of perturbation theory, is also described
and suggests new possibilities for the underlying dynamics.

We have surveyed the possibility that
the observed Higgs boson with a Coleman-Weinberg potential
is {\em maximally observable} at the LHC. 
To achieve this
we assume a minimal extension of the standard model
consisting of a second, ``dormant,'' Higgs doublet that couples to the
standard model $SU(2)\times U(1)$ gauge fields.
The dormant Higgs doublet can sculpt 
a Coleman-Weinberg potential for the Higgs boson provided
it has a   mass of about $\sim 380\pm 10\% $ GeV.

The new doublet, coupled
to standard model $SU(2)\times U(1)$, is pair produced at the LHC
in  $pp\rightarrow \gamma, Z^0\rightarrow 
H^+H^-, H^0 A^0$ and $pp\rightarrow W^\pm \rightarrow 
H^\pm +(A^0, H^0)$  at the $\sim 1 $ fb level. It can
naturally couple strongly to some SM fermions, and we
consider the case of $O(1)$ coupling to the b-quark. In this
case the production can be via gluon fusion with $\sim 10$ to $100$ fb
cross-sections. The coupling to the fermions, albeit
model dependent, is essential to make visible final states at the LHC.
In the cases considered we are encouraged that the new states
may be observable in Run-II at the LHC.

The departures from the standard model Higgs potential, the trilinear, quadrilinear
and even quintic self-couplings, are fairly significantly modified
in this scenario, and may also be
addressable at the LHC, and certainly at future Higgs factories. 

We have also described a schematic model in which the 
CW potential arises at the two-loop level via new fermions.
These would have masses at the order of $\sim 230$ GeV, and would be pair
produced at the LHC. We will develop this idea further elsewhere.

The general idea that ``the Higgs mass  comes from quantum mechanics'' is, to us,
sufficiently compelling to warrant the present   phenomenological approach and ask if
there is any evidence, potentially visible to experiment,  that can
determine whether the CW mechanism is operant for the Higgs boson. 
As such, our focus has presently largely left
the UV completion issues untouched.

Yes, there are certainly challenges and difficulties in constructing a 
UV complete scenario  (see, e.g., \cite{LR,schmalz}).
The main problem with our simple phenomenological model is the
occurance of nearby Landau poles in the running quartic couplings,
that are reached at $\sim 10$ TeV.  These are either
blemishes on the scheme, or may be harbingers of new physics, such as
compositeness of the new bosonic states, \cite{Miransky,BHL,Dobrescu,Fukano}.  We've 
only briefly discussed UV completion issues, as we feel these issues are secondary.
We plan to return to these issues in greater detail elsewhere \cite{Ross}.

If we could establish  a Coleman-Weinberg origin of
the Higgs boson mass, we would then have two scales in nature generated
by quantum loop effects: $\Lambda_{QCD}$ and $v_{weak}$. 
The grand hypothesis that: ``all mass in nature comes from quantum mechanics''
would gain significant validation. Our view of the UV would then
have to accomodate it.  This hypothesis 
may ultimately imply a radically different view of nature than 
our current ``GUTs to strings'' philosophy.  

Some of these issues and  ``predictions'' have
been discussed elsewhere \cite{Hill10}.  For example,  
we live in a $D=4$ universe, and 
it is striking that $D=4$ is
the only possibility for classical scale symmetry given
Yang-Mills field theories as an underpinning of nature, since the 
trace of the Yang-Mills field stress tensor is classically zero only in $D=4$. 
Quantum mechanics then supplies the trace anomaly and allows for the generation
of mass and large hierarchies through the renormalization group. 
We see this with QCD and the compelling question is whether it
also applies to the weak scale and Higgs boson.  Hence,
the hypothesis that ``all mass in nature comes from quantum mechanics''
already seems broadly consistent with our $D=4$, large universe.
The stakes are high: this may ultimately require a classically scale-invariant approach to gravity,
such as $D=4$ Weyl gravity with a quantum, QCD-like origin of $M_{Planck}$
\cite{adler} (see also \cite{Hill10} and references therein).

\vskip 0.5in
\noindent
{\bf Acknowledgements}
\vskip .2in
I wish to thank for useful discussions, 
K. Allison, W. Bardeen, E. Eichten, H. Frisch,  G. G. Ross, P. Steinhardt, R. Vidal, C. Wagner, 
and various members of
the ATLAS collaboration, particularly J. Allison and D. Miller. 

\newpage

\vskip .1in
\noindent
{\bf Appendix A: Scalar Field Stress Tensors}
\vskip .1in
\renewcommand{\theequation}{A.\arabic{equation}}   
\setcounter{equation}{0}  % reset counter 

Consider a scalar field theory in flat spacetime with Minkowski metric $
\eta _{\mu \nu }$:
\beq
\label{theory}
S=\int d^{4}x\;{\cal{L}}=\int d^{4}x\left( \frac{1}{2}\partial _{\mu }\phi \partial
^{\mu }\phi - V(\phi)\right) 
\eeq
where the Euler-Lagrange equation of motion is:
\beq
\partial ^{\mu }\frac{\delta S}{\delta \partial _{\mu }\phi }-\frac{\delta S%
}{\delta \phi }=\partial ^{2}\phi + \frac{\delta}{\delta\phi} V(\phi )=0
\eeq
% and $V'(\phi) = \frac{\delta}{\delta\phi} V(\phi )$.
We perform an infinitesimal diffeomorphism in the flat space theory
holding the metric fixed:
\beq
x^{\mu \prime }=x^{\mu }-\zeta ^{\mu }(x)
\eeq 
where the scalar field is invariant,
$\phi ^{\prime }(x^{\prime })=\phi (x)$
\footnote{Note that here 
we can define $\phi'(x') = \phi(x') + \delta\phi(x)
= \phi(x)+\zeta^\mu \partial_\mu \phi(x) +\delta\phi(x) $
and hence $
\delta\phi(x) = -\zeta^\mu \partial_\mu \phi(x) $, and
no additional terms are generated; alternatively
we can do an {\em active tranformation} 
$\phi(x)\rightarrow \phi(x)  +\zeta^\mu \partial_\mu \phi(x)$
and additional terms are generated but vanish by 
integration by parts and use of  equations of motion.},
but the coordinate differentials transform as:
\bea
\delta dx^{\mu } & = & -d\zeta ^{\mu }(x) = -\left( \partial
_{\lambda }\zeta ^{\mu }\right) dx^{\lambda }
\nonumber \\
\delta \partial _{\mu }^{ }& = & (\partial ^{\nu }\zeta _{\mu
})\partial _{\nu }
\nonumber \\
\delta d^{4}x^{ } & = & -(\partial _{\mu }\zeta ^{\mu })d^{4}x
\eea
The action transforms as:
\bea
\delta S &  = & \int d^{4}x[-\frac{1}{2}(\partial _{\rho }\zeta ^{\rho })\partial
_{\mu }\phi \partial ^{\mu }\phi +(\partial ^{\rho }\zeta _{\mu })\partial
_{\rho }\phi \partial ^{\mu }\phi 
\nonumber \\
& & 
\qquad +(\partial _{\mu }\zeta ^{\mu
})V(\phi )]
\nonumber \\
\qquad & \equiv & \half\int d^{4}x\left[ \left( \partial _{\mu }\zeta _{\nu }
+\partial _{\nu }\zeta _{\mu } \right)
T^{\mu \nu }\right] 
\eea
and the resulting {\em canonical stress tensor} is:
\bea
T_{\mu \nu } & = &  \partial _{\mu }\phi \partial _{\nu
}\phi -\eta _{\mu \nu }\left( \frac{1}{2}\partial _{\rho }\phi \partial
^{\rho }\phi -V(\phi) \right)
\eea
Note the divergence of the stress tensor:
\bea
\partial ^{\mu }T_{\mu \nu } & = & \partial ^{2}\phi \partial _{\nu }\phi
+\partial _{\mu }\phi \partial ^{\mu }\partial _{\nu }\phi -
\partial _{\nu }\left( \frac{1}{2} \partial _{\mu }\phi \partial ^{\mu }\phi -
V(\phi) \right) 
\nonumber \\
 & = & \partial _{\nu }\phi \left(
\partial ^{2}\phi +V'(\phi)\right) 
\eea
The stress tensor is the Noether current associated with translations 
in space and time. The
conservation of the stress tensor is a consequence
of these symmetries and implies the equation of motion.

We can choose, however,  $\zeta ^{\mu }=-\epsilon x^\mu$,
corresponding to an infinitesimal scale transformation.
The action then varies as:
\bea
\delta S &  = &  \int d^{4}x\left[ \left( \partial _{\mu }\epsilon x_{\nu }\right)
T^{\mu \nu }\right] 
\eea
and the scale current is defined by:
\beq
\frac{\delta S}{\partial_\mu \epsilon} \equiv S^\mu =  x_{\nu }T^{\mu \nu }
\eeq
with divergence:
\beq
\partial_\mu S^\mu = T^{\mu}_{ \mu }
\eeq
The canonical stress tensor, however, has a nonzero trace, even 
when $V(\phi)$ is scale invariant. 
\beq
T_{\mu }^{\mu }=-\partial _{\rho }\phi \partial ^{\rho }\phi +4V(\phi)
\eeq
It can be ``improved'' to yield a vanishing trace in the
scale invariant case, \eg, when $V(\phi) \propto \phi^4$ \cite{CCJ}.

\vskip .2in
\noindent
{\bf Stress Tensor Improvement}
\vskip .1in

We add to the action a total divergence:
\beq
S \rightarrow S + S_2 \qquad S_{2}= \int d^{4}x \; \xi \partial ^{2}\phi ^{2}
\eeq
where $\xi(x)$ can be viewed as an arbitrary function
of space-time, but we take the limit $\xi\rightarrow\xi_0$ (constant) after
manipulating the action. With constant $\xi$
this is  a surface term and
does not affect the equations of motion.  However,
it varies under the  diffeomorphism to produce a nonvanishing result:
\bea
\delta S_{2}  & =  & \xi \int d^{4}x [-(\partial _{\mu }\zeta ^{\mu })\partial
^{2}\phi ^{2}+\partial ^{\mu }((\partial ^{\nu }\zeta _{\mu })\partial _{\nu
}\phi ^{2})  \nonumber \\
& & \qquad +(\partial ^{\nu }\zeta _{\mu })\partial _{\nu }\partial ^{\mu
}\phi ^{2}] + O(\partial\xi)
\eea
Note that the second term in the $\xi\rightarrow$ (constant) limit 
is an irrelevant surface term, but the first and third terms yield:
\bea
\delta S_{2}  
& =  &  -\xi_0 \int d^{4}x(\partial _{\mu }\zeta _{\nu })[\eta ^{\mu \nu
}\partial ^{2}\phi ^{2}-\partial ^{\nu }\partial ^{\mu }\phi ^{2}]
\nonumber \\
 & \equiv & \half \int d^{4}x(\partial _{\mu }\zeta _{\nu }+ \partial _{\nu }\zeta _{\mu })[Q^{\mu \nu }]
\eea
where:
\bea
Q_{\mu \nu } & = & \xi_0 (\partial _{\mu }\partial _{\upsilon }\phi ^{2}-\eta _{\mu
\nu }\partial ^{2}\phi ^{2})
\eea
$Q_{\mu \nu }$ has the trace: 
\bea
\!\!\!\!\!  \!\!\!\!\!  
Q_{\mu }^{\mu } & = & -3\xi_0 (\partial
^{2}\phi ^{2})=-6\xi_0 (\phi \partial ^{2}\phi +\partial _{\rho }\phi \partial
^{\rho }\phi )
\eea
We thus choose $\xi_0 =-\frac{1}{6}$ and obtain the
``improved stress tensor'':
\bea
\label{improved}
\widetilde{T}_{\mu \nu } & = & T_{\mu \nu }+Q_{\mu \nu }
\nonumber \\
& = &\frac{2}{3} \partial
_{\mu }\phi \partial_{\nu }\phi -\frac{1}{6}\eta _{\mu \nu }\partial _{\rho
}\phi \partial ^{\rho }\phi -\frac{1}{3}\phi \partial _{\mu }\partial
_{\upsilon }\phi 
\nonumber \\
& & 
\!\!\!\!\!
+\;\frac{1}{3}\eta _{\mu \nu }\phi \partial ^{2}\phi +
\eta _{\mu \nu }V(\phi)
\eea
The conservation of the stress tensor is unaffected by adding
the conserved improvement term $Q_{\mu\nu}$. 
However, we see that the trace now yields:
\beq
\label{trace}
\widetilde{T}_{\mu }^{\mu }=\phi \partial ^{2}\phi
+ 4V(\phi) = -\phi \frac{\delta}{\delta \phi} V(\phi) + 4 V(\phi) 
\eeq

We can also generate the improved stress tensor
by including the``conformal coupling'' of the scalar field to gravity, in the
action:
\beq
S=\frac{1}{2}\int \sqrt{-g}d^{4}x\left( g_{\mu \nu }\partial ^{\mu }\phi
\partial ^{\nu }\phi -V(\phi) - \xi_0 R\phi ^{2}\right) 
\eeq
In weak field gravity 
the metric is expanded about the flat Minkowski metric:  
\beq
g_{\mu \nu } =\eta _{\mu \nu }+h_{\mu \nu }
\qquad
 g^{\mu \nu }=\eta ^{\mu \nu }-h^{\mu \nu }
\eeq
and to $O(h_{\mu \nu })$:
\bea 
R & = & \partial ^{2}h-\partial ^{\mu }\partial ^{\nu }h_{\mu \nu }
\nonumber \\
\sqrt{-g} & = & 1 +\half h
\eea
where: $\eta ^{\mu \nu }h_{\mu \nu }\equiv h$ (signs are tricky here).

We choose $\xi_0 =\frac{1}{6}$ and the first order action becomes:
\bea
S & = & \frac{1}{2}\int d^{4}x[\eta ^{\mu \nu }\partial _{\mu }\phi \partial_{\nu }\phi 
-V(\phi)-h^{\mu \nu }\partial _{\mu }\phi \partial _{\nu}\phi 
\nonumber \\
& & \!\!\!\!\! \!\!\!\!\! \!\!\!\!\! \!\!\!\!\! 
 +\;\frac{1}{2}\eta ^{\mu \nu }h_{\mu \nu }\left( \partial _{\rho }\phi\partial ^{\rho }\phi -V(\phi)
\right) -\frac{1}{6}\left( \partial ^{2}h-\partial ^{\mu }\partial ^{\nu }h_{\mu \nu}\right) \phi ^{2}]
\nonumber \\
& = & S_{0}-\frac{1}{2} \int d^{4}x\; h^{\mu \nu }\widetilde{T}_{\mu \nu }
\eea
Hence, a small variation in the metric about flat spacetime,
$\delta g_{\mu\nu} = h_{\mu\nu}$, generates the improved stress tensor
with the inclusion of the conformal term. 

Note that the $\xi\partial^2\phi^2$ term does not affect the local metric
variation in curved space since:
\beq
\int d^4x\;\sqrt{-g} D_\mu \partial^\mu \phi^2 = \int d^4x\; \partial_\mu (\sqrt{-g} \partial^\mu \phi^2 ) 
\eeq
where $D_\mu$ is the covariant derivative. We see that this
is a surface term and is insensitive to a local variation $\delta g_{\mu\nu}$.

We've seen that the variation of the action in flat space by the
diffeomorphism,  $\delta x^{\mu }=\zeta ^{\mu }$ generates
the improved stress tensor $\widetilde{T}_{\mu \nu }$ in the presence of the
 $\xi\partial^2\phi^2$ term. Likewise a variation of the
metric generates the improvement with the conformal coupling term.
An Einstein transformation (general covariance) implies:
\beq
\label{GI}
\delta g_{\mu\nu} = h_{\mu\nu} = \partial_\mu\zeta_\nu + \partial_\nu\zeta_\mu
\eeq
If we perform both of these
transformations together we obtain:
\bea
\label{diffgen}
\delta S &  = &  \half \int d^{4}x\left[ \left( \partial _{\mu }\zeta _{\nu } 
+   \partial _{\nu }\zeta _{\mu }- h_{\mu\nu}\right)
\widehat{T}^{\mu \nu }\right]
\eea
which is zero when eq.(\ref{GI}) is applied. This
is now a gauge transformation. 
The diffeomorphism on the ``matter side'' cancels
the variation wrt the metric on the ``gravity side'', and
both transformations generate a conserved improved stress tensor.
This is analogous to any gauge theory, such as QED, where
we can generate the current by doing a local gauge transformation
of the electron wave-function ($\sim $ ``matter side'') or
by varying the action wrt the vector potential  ($\sim $ ``gravity side'').
We can  define the Noether current for scale symmetry by either procedure.

\newpage
\vskip .1in
\noindent
{\bf Appendix B.  Trace Anomaly and Feynman Loops}
\vskip .1in
\renewcommand{\theequation}{B.\arabic{equation}}   
\setcounter{equation}{0}  % reset counter 

A classically scale invariant potential is defined by
the condition:
\beq
\phi \frac{\delta}{\delta \phi} V(\phi) = D V(\phi) \;\;\makebox{where $D=4$.}
\eeq
 For a classically scale invariant potential the improved
stress tensor trace, eq.(\ref{trace}), vanishes, 
and the associated scale current is conserved.

In general, $D= 4 +\gamma$ where $\gamma$ is the ``anomalous dimension''
of the potential.  Such is the case for Coleman-Weinberg
potentials where the running of the coupling is included.  For
example, if we choose,
\beq
V(\phi) = \frac{\lambda(\phi)}{4} \phi^4, \qquad
\makebox{and} \qquad \phi \frac{\delta}{\delta \phi} \lambda(\phi) = \beta(\lambda)
\eeq
then we see:
\beq
\label{traceanom}
\widetilde{T}_{\mu }^{\mu }= -\phi\frac{\delta}{\delta \phi} V(\phi) + 4V(\phi) 
= -\frac{\beta(\lambda)}{\lambda} V(\phi) 
\eeq
This is called the trace-anomaly; $\gamma = \beta/\lambda $ is the anomalous dimension.

\begin{figure}[t]
\vspace{5cm}
\includegraphics{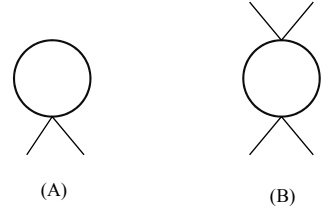}
\vspace{-1.0cm}
\caption[]{ Diagrams for counterterms (A) $\delta m^2$; (B) $\delta \lambda$.}
\label{cts}
\end{figure}

Let us examine how the trace anomaly arises at the one-loop  level
via a direct calculation of the effective potential.  Consider  the 
real scalar field theory lagrangian:
\beq
\label{S0}
 L =  \half (\partial\phi)^2 -\half  m^2 \phi^2 - \frac{1}{4} \lambda \phi^4 
\eeq
We define renormalized couplings and
 $O(\hbar)$ counterterms:
\bea
\label{ct0}
m^2 &  = & m_r^2 + [\hbar] \delta m^2 
\nonumber \\
\lambda & = & \lambda_r + [\hbar] \delta\lambda 
\eea
%[Note scalar propagator = +i/(p^2-m^2)
The counterterms can be
computed from the 1PI scattering
amplitudes of Figs.(\ref{cts}A, \ref{cts}B). 
%In the background
%field loop expansion this is equivalent to:
%\bea
%\label{ct}
%0 & = & -\frac{i}{2}\delta m^2\phi_c^2 - \frac{3i\lambda\phi_c^2}{2}\int \frac{d^4 \ell}{(2\pi)^4}\frac{i}%{\ell^2 -m_r^2}%
%\nonumber \\
%0 & = & -\frac{i}{4}\delta \lambda \phi_c^4 + \frac{9\lambda^2\phi_c^4}{4}\int \frac{d^4 \ell}{(2\pi)^4}\frac%{1}{(\ell^2 - m_r^2)^2}
%\eea
We obtain:
\bea
\label{ct2}
\delta m^2 &= & -\frac{3\lambda}{16\pi^2}\left(\Lambda^2 -m_r^2\ln(\Lambda^2/\mu^2)\right)
\nonumber  \\ 
\delta \lambda  &= & \frac{9\lambda^2}{ 16\pi^2}(\log(\Lambda^2/m_r^2) - 1)
\eea
Here we  define the Feynman
loops with Euclidean momentum
space cut-off and neglect external momenta
in the loops. There is no wave-function renormalization
constant as there is no external momentum flow through
the loop of Fig.(\ref{cts}A) \ie, the theory is super-renormalizeable.

We use the $\hbar$ expansion
and work in a classical background field, $\phi_c$.
We  introduce a classical source term in the lagrangian, $- J\phi$. 
This induces the shift in the field, 
\beq
\phi = \phi_c + \hbar^{1/2}\hat{\phi}
\eeq
where $\phi_c$ satisfies the renormalized equation of
motion, $\partial^2\phi_c + m_r^2\phi_c + \lambda_r\phi_c^3 + J = 0$
\footnote{The reason for introducing the source term is to remove all the linear 
cross-terms,
$\propto \hat{\phi}$, arising from the shift. In this perturbative approach
there remain $O(\hbar^{3/2})$ terms, $\sim \delta m^2\phi_c\hat{\phi}$.
These we ignore since we are working to $O(\hbar)$.
We then add back a term $+ J\phi_c $ which cancels the
$- J\phi_c $ arising from the shift. The general formalism
of the Legendre transformed potential is given in ref.(\cite{Coleman:1973jx}). }.
The lagrangian becomes:
\bea
L = L_0(\phi_c) + [\hbar]\hat{L}(\phi_c, \phi)
\eea
where, to $O(\hbar)$:
\beq
\label{Sc}
 L_0(\phi_c) =  \half (\partial\phi_c)^2 -\half  m_r^2 \phi_c^2 - \frac{1}{4} \lambda_r \phi_c^4 
\eeq
and:
\bea
\hat{L}(\phi_c, \phi) & = &  \half (\partial\hat\phi)^2 -\half  (m^2  + 3\lambda\phi_c^2)\hat\phi^2 
+ ...
\nonumber \\
& & \!\!\!\!\!  -\half  \delta m^2\phi_c^2 - \frac{1}{4}\delta  \lambda \phi_c^4 
\eea
where the $+...$ refers to terms of higher order in $\hbar$.

%\begin{figure}[h]
%\includegraphics[bb = 0 0 100 100 ]{UV1.pdf}
%\caption{UV running of the Dormant Higgs model}
%\end{figure}

We now integrate out the quantum fluctuations, $\hat\phi$. 
Let $Z=\int D\hat \phi \exp(i\int d^4x \hat{L}/\hbar)$ be the path integral. The effective lagrangian becomes $L_0 -i\hbar\ln(Z)$, which takes the form:
\bea
\label{Leff}
L_{eff} & = & L_0(\phi_c) + i\hbar  \int \frac{d^4\ell}{(2\pi)^4}\ln(\ell^2 -m_r^2 -3\lambda_r\phi_c^2)
\nonumber \\
& &  -\half  \delta m^2\phi_c^2 - \frac{1}{4}\delta  \lambda \phi_c^4 
\eea
Note that the second term acquires a  sign flip since 
$ -i\hbar\ln(Z)\sim -i\hbar \int \ln(1/(\ell^2 - m^2))\sim i\hbar \int \ln(\ell^2 - m^2)$. We  drop 
irrelevant additive constants. 

The integral can be done by performing a Wick rotation 
($\ell_0 \rightarrow i \ell_0$, $\ell^2 \rightarrow -\ell_0^2 -\vec{\ell}^2$,
and $d^4\ell \rightarrow i d\ell_0 d^3\ell$ ) and we use a 
Euclidean momentum space cut-off, $\Lambda$.
Up to additive constants $\propto \Lambda^4, m_r^2\Lambda^2 $, we obtain:
\bea
& &  i\hbar  \int \frac{d^4\ell}{(2\pi)^4}\ln(\ell^2 -m_r^2 -3\lambda_r\phi_c^2) =
 -\frac{1}{32\pi^2}\times  
\nonumber \\
& & 
%\!\!\!\!\!  \!\!
\left[  3\lambda\phi_c^2 \Lambda^2 
-\half (m_r^2 +3\lambda\phi_c^2 )^2\left(\ln\left(\frac{\Lambda^2}{m_r^2 + 3\lambda \phi_c^2}\right)-\half\right) \right]
\nonumber \\
& & 
\eea
If we now add in the counterterms of eq.(\ref{ct2}) as in eq.(\ref{Leff}) we have:
\bea
\label{Leff2}
L_{eff} & = & \half (\partial\phi_c)^2   - V_{eff}
\eea
where the effective potential is:
\bea
& & V_{eff}  = V_0(\phi_c)
\nonumber \\
& & 
+ \frac{1}{32\pi^2}\left(  3\lambda m_r^2\phi_c^2 +\half m_r^4 + \half 9\lambda^2\phi_c^4
\right) \ln\left( 1+ \frac{  3\lambda \phi_c^2 }{m_r^2 } \right)
\nonumber \\
& & 
-\frac{1}{64\pi^2 }\left(  3\lambda m_r^2\phi_c^2 + \half 9\lambda^2\phi_c^4  \right)
\eea
where:
\beq
V_0 = \half  m_r^2 \phi_c^2+\frac{1}{4} \lambda_r \phi_c^4
\eeq
%Lagrangian, not potential.
The classically scale invariant limit corresponds to $m_r^2 \rightarrow 0$.
The potential then becomes:
\bea
\label{Veff0}
 \!\!\!\!\!
V_{eff} & = & \frac{1}{4} \lambda_r \phi_c^4
 + \frac{ 9\lambda^2\phi_c^4}{64\pi^2}
\left[ \ln\left( 1+ \frac{  3\lambda \phi_c^2 }{m_r^2 } \right) -\half \right]
\nonumber \\
& &
\eea
Note that there is an infrared divergence in the limit $m_r^2 \rightarrow 0$
and we retain the $m_r^2$ in the argument of the log as an infrared regulator.

We can now constuct the improved stress tensor from the full effective
lagrangian eq.(\ref{Leff2}):
\bea
\label{improvedc}
\widetilde{T}_{\mu \nu } & = & \frac{2}{3} \partial
_{\mu }\phi_c \partial_{\nu }\phi_c -\frac{1}{6}\eta _{\mu \nu }\partial _{\rho
}\phi_c \partial ^{\rho }\phi_c -\frac{1}{3}\phi_c \partial _{\mu }\partial
_{\upsilon }\phi_c 
\nonumber \\
& & 
\!\!\!\!\!
+\;\frac{1}{3}\eta _{\mu \nu }\phi_c \partial ^{2}\phi_c +
\eta _{\mu \nu }V_{eff}(\phi) 
\eea
The equation of motion of $\phi_c$ is now $\partial_\mu \hat{T}^\mu_\nu = 0$
and includes the $O(\hbar)$ quantum effects,
\beq
 0 = \partial^2\phi_c + \frac{\delta }{\delta \phi_c}V_{eff}(\phi_c)
\eeq
The trace of the improved stress tensor  is therefore:
\bea
\widetilde{T}_{\mu }^{\mu }(\phi_c) & = &-\phi_c\frac{\delta}{\delta \phi_c} V_{eff}(\phi_c) + 4V_{eff}(\phi_c)
=  -\frac{ 9\lambda^2\phi_c^4}{32\pi^2}
\nonumber \\
& &
\eea
where the latter term arises from the derivative of the logarithm
in eq.(\ref{Veff0}).

Note that we can infer the $\beta$-function of $\lambda$ 
from $\delta\lambda$ in eq.(\ref{ct2}).
With our sign convention, $\lambda_r = \lambda -\delta\lambda$,
and we can identify the ``running RG scale'' $\mu$ with
$m_r$, hence :
\beq
\frac{d\lambda_r}{d\ln(\mu)}= \frac{9\lambda^2}{8\pi^2}
\eeq
Comparing expressions
we thus see that the trace anomaly is:
\beq
\widetilde{T}_{\mu }^{\mu } = -\frac{1}{4}\beta(\lambda)\phi_c^4 = -\frac{\beta}{\lambda }V_0(\phi_c)
\eeq
We therefore observe that, for a theory with vanishing
renormalized mass, $m^2_r = 0$, we have a violation of scale symmetry
by the trace anomaly, $\propto \beta(\lambda)$, which is $O(\hbar)$ and
represents the RG running ot $\lambda$.
We have no other such source of scale violation in this limit.  
There are, of course,  infrared singularities
in higher order terms in the expansion, $\phi_c^{2N}/\mu^{2N}$ for $N>2$, but
these are associated with the long-distance physics
of the matrix element of the trace anomaly operator itself.  

The important implication of this result, as emphasized by Bardeen, \cite{bardeen2,bardeen}, 
is that the additive quadratic divergence needed to renormalize the mass is an artifact of
our calculational procedure and has nothing to do with the physics of mass generation.
The scale current and its divergence controls the physics of mass generation.
A scale invariant field theory is one whose scale current is strictly conserved
to all orders in perturbation theory; a
classically scale invariant theory will typically experience scale breaking by the trace anomaly,
but the additive quadratic divergence, $\Lambda ^2$, encountered in the 
calculational procedure is a red herring.

\vskip .3in
\noindent
{\bf  Appendix C: Classic Coleman-Weinberg Potentials from the Renormalization Group}
\vskip .2in
\renewcommand{\theequation}{C.\arabic{equation}}   
\setcounter{equation}{0}  % reset counter 

There are various renormalization groups.  The 
relevant RG depends upon the application. For example,
running a set of coupling constants in external momenta
for scattering amplitudes, such as the QCD
coupling $g_3$ with a single scale $\mu$, is a typical application. Particles such
as the top quark then decouple at $\mu \sim m_t$, and the $\beta$-function counts
only the active light quarks below that scale, and this affects the evolution
of $g_3$ and contributes to the value of $\Lambda_{QCD}$. 

In applications to the CW potential we are interested in running of coupling constants
where the scale $\mu$ is replaced by the  field,  $\phi$, itself.
Here we are only including the low momentum components of $\phi$, in particular the
zero-momentum VEV of $\phi$.  If $\phi$ is the standard model
Higgs boson, with its classical mass term set to zero, we want to vary
$\phi$ over a large range of scales to find a minimum of
the effective potential.  Since the top quark receives its mass from this VEV, then {\em the top quark never
decouples as we run $\phi$ to lower values}.  The same is true for any field, such as $H_2$,
that receives its mass from the VEV of $\phi$.

This is a surprising and counter-intuitive
effect: if we were to run $\phi$ down to the QCD scale, for example, the QCD coupling would run with
$\phi$ as well, but the top quark (and $b$ and $c$ quarks as well) would remain active 
far into the infrared.  This has the stunning effect of reducing $\Lambda_{QCD}$ to about half of its normal value.  
Of course, the QCD chiral phase transition would still occur, at about $\sim 500 $ GeV, and the
resulting $SU(6)\times SU(6)\rightarrow SU(6)$  chiral breaking would occur, with 35 Nambu-Goldstone
pions, and a QCD constituent quark mass would be generated for all 6 quarks. 
The Higgs doublet has the control VEV, $\phi$, and this yields three additional massles NGB's. 
These would mix with some of the pions, three of which would be eaten to break the 
$SU(2)\times U(1)$ electroweak symmetry.  This is a dynamically stable
``minimum of the Higgs effective potential'' and it is
similar to the way in which QCD would act as technicolor 
if there were no Higgs boson in the standard model. 

Let's consider in greater detail the direct derivation of Coleman-Weinberg potentials
using the renormalization group.  
As an exercise, comparing to the derivation
of  ref.\cite{Coleman:1973jx}, we'll use
the RG method to derive the potential for massless scalar electrodynamics:
\beq
|(i\partial _{\mu }-eA_{\mu })\phi |^{2}-\frac{\lambda }{2}|\phi |^{4}
\eeq
The RG equation for the quartic
coupling is:
\beq
\beta(\lambda,e) = \frac{d\lambda }{d\ln (\mu )}=\frac{1}{16\pi ^{2}}\left( 10\lambda ^{2}-12\lambda
e^{2}+12e^{4}\right) 
\eeq
Note that this is similar to the RG equation for 
$\lambda_{1}$
for a single Higgs boson in the standard model, as in \cite{HLR}, eq.(6a). The term, 
$ 12\lambda_{1}^{2}/{16\pi^{2}} $ 
has become $ 10\lambda _{1}^{2}/{16\pi^{2}}$ since
the coefficient is $\propto 8+2N$ where
$N=2$ for a Higgs doublet and $N=1$ for a complex singlet $\phi $,
and $N=1/2$ for the real scalar field as we discussed above. The other
terms are obtained by setting $g_{2}=0$ and $\frac{1}{2}g_{1}=e$ where
the  $\frac{1}{2}$ factor is the weak hypercharge.

Consider the classical effective potential:
\beq
V(\phi )=\frac{\lambda \left( |\phi |\right) }{2}|\phi |^{4}
\eeq
and we again obtain $\lambda \left( |\phi |\right) $ by solving the RG
equation.
We thus obtain in leading order where on the {\em rhs} $\lambda$ and $e$ are
approximated as constants:
\beq
\label{qed}
V(\phi )=\frac{\lambda _{0}}{2}|\phi |^{4}+\frac{1}{16\pi ^{2}}\left(
5\lambda ^{2}-6\lambda e^{2}+6e^{4}\right) |\phi |^{4}\ln \left(\frac{|\phi |}{M}\right)
\eeq
To compare to ref.\cite{Coleman:1973jx}, eq.(4.5), we note the CW
normalization conventions, 
\beq
 \phi_{c}^{2}=2|\phi |^{2} \qquad \makebox{and} \qquad
 \frac{\lambda _{CW}}{4!}\phi _{c}^{4}=\frac{\lambda _{0}}{2}|\phi |^{4}
\eeq
thus ${\lambda _{CW}}=3\lambda _{0}$. We are consistent
in the $\lambda^2$ and $e^4$ terms with their result, ref.\cite{Coleman:1973jx} eq.(4.5):
\beq
\label{IIone}
V(\phi _{c}^{\prime })=\frac{\lambda _{CW}}{4!}\phi _{c}^{\prime 4}+\left( 
\frac{5\lambda _{CW}^{2}}{1152\pi ^{2}}+\frac{3e^{4}}{64\pi ^{2}}\right) 
\phi _{c}^{\prime 4}\ln \left(\frac{\phi
_{c}^{2}}{M^{\prime 2}}\right)
\eeq
where $M^{\prime 2}=2M$
%\bea
%V(\phi _{c}) & = &\frac{\lambda _{CW}}{4!}\phi _{c}^{4}
%\nonumber \\
%& & \!\!\!\!\!\!\!\!   \!\!\!\!\!\!\!\!   \!\!\!\!\!\!\!\!   \!\!\!\!\!\!\!\!   
%+\left( \frac{5\lambda _{CW}^{2}}{%
%1152\pi ^{2}}-\frac{e^{2}\lambda _{CW}}{64\pi ^{2}}+\frac{3e^{4}
%}{64\pi ^{2}}\right) \ \phi _{c}^{4}\ln \left(\frac{\phi^2_{c}}{M^{\prime 2}}\right)
%\eea

We see one discrepancy in
the presence of the $e^{2}\lambda$ 
term in eq.(\ref{qed}) which is absent in eq.(\ref{IIone}). 
The $e^2 \lambda$ term arises for us because we have enforced the 
canonical wave-function normalization (kinetic term normalization) in our
definition of $\phi _{c}$ \ie, canonical wave-function 
normalization is implict in our RG equation.
To this order, however, we can absorb away the $e^2\lambda$
term by a field redefinition
and it therefore does not affect the 
potential.
We then identically reproduce the exact form of CW eq.(\ref{IIone}).

We also see, however,  that the $\lambda^2$ term
is irrelevant
since we can absorb an additional $\lambda$ factor into $\phi$.
With the net redefinition:
\beq
\phi=\phi^{\prime }\left[1+\left(\frac{6}{16\pi ^{2}}e^{2}-\frac{5}{16\pi ^{2}}\lambda_{CW}
\right)\ln \left(\frac{\phi
}{M}\right)\right]
\eeq
the resulting effective potential then contains only
two relevant terms, the classical quartic coupling
and the $O(e^4)$ interaction term:
\beq
\label{qed2}
V(\phi )=\frac{\lambda _{0}}{2}|\phi |^{4}+\frac{3e^{4}}{16\pi ^{2}} |\phi |^{4}\ln \left(\frac{|\phi |^2}{M^2}\right)
\eeq
Indeed, as discussed by ref.\cite{Coleman:1973jx},
the only possible non-trivial perturbative minima of the effective potential
 involves exclusively these two terms. 
Moreover, the rescaled RG equation takes the form:
\beq
\beta'(\lambda,e) = \frac{d\lambda' }{d\ln (\mu' )}=\frac{12e^{4}}{16\pi^{2}} 
\eeq

We also see that the RG admits a solution in which $\lambda(\phi)$
can be negative and cross to positive
values with positive $\beta$.  
The Landau pole occurs at $\phi_L$ and is 
is determined by the condition that the wave-function
of $\phi$ is vanishing:
\beq
Z'(\phi_L) = 0 =\left[1+\left(\frac{6}{16\pi ^{2}}e^{2}-\frac{5}{16\pi ^{2}}\lambda_{CW}
\right)\ln \left(\frac{\phi_L}{M}\right)\right]
\eeq

This form of the potential makes contact with the 
functional integral calculation where the photon mass is $M^2_\gamma=e^2|v|^2$:
\beq
\tilde{V}(\phi _{c}^{\prime })=
\sum_i\frac{ M^4_{\gamma i}}{64\pi ^{2}}
(\phi _{c}^{\prime 4}/v^4)\ln \left(\frac{\phi
_{c}^{2}}{v^{\prime 2}e^{-1/4}}\right)
\eeq
where the sum counts the $3$ spin states of the photon.  Note that the
massive $\phi$ contribution is not counted in this normalization. 
If we had not absorbed away the $\lambda^2$ term we would find a mismatch
in the coefficient of the $m_\phi^4$ term with the usual log path integral
result.  The RG equation is counting degrees of freedom in the
symmetric phase, while the $\ln \det(\partial^2 + m^2)$ result counts
only the real scalar (``Higgs'') degree of freedom and 
not the eaten Nambu-Goldstone modes (which are counted in the factor of 3 for
the massive photon).  The mismatch is present
in general in this term, but is irrelevant for perturbative Coleman-Weinberg
potentials.

We finally remark that for applications to dynamical situations, such as slow-roll
inflationary models, it would be a blunder 
to ignore the wave-function renormalization terms, and one should adopt the
canonically normalized form of the potential as in eq.(\ref{qed}).
The slow-roll physical field motion is defined by the canonical
normalization, so predictions of obervables may depend upon maintaining the
canonical normalization.

\newpage
%\vskip .3in
\noindent
{\bf  Appendix D: Quintic Order terms in the Coleman-Weinberg Potential
}
\vskip .2in
\renewcommand{\theequation}{D.\arabic{equation}}   
\setcounter{equation}{0}  % reset counter 

The fifth derivative of the quartic coupling is:
\bea
v^5\frac{d^5 \lambda_1}{dv^5} & = & 
\beta _{i}\beta _{j}\beta _{k}\beta _{\ell }\frac{\partial^{4}\beta }{\partial \lambda _{i}\partial \lambda_{j}\partial \lambda_{k}\partial \lambda _{\ell }}
\nonumber \\
& & \!\!\!\!\!  \!\!\!\!\!  \!\!\!\!\!  \!\!\!\!\! \!\!\!\!\!  \!\!\!\!\!  
+ \beta _{\ell }\frac{\partial \beta _{k}}{%
\partial \lambda _{\ell }}\frac{\partial \beta _{j}}{\partial \lambda _{k}}
\frac{\partial \beta _{i}}{\partial \lambda _{j}}\frac{\partial \beta }{
\partial \lambda _{i}}
+
6\beta _{i}\beta _{j}\beta _{k}\frac{\partial \beta
_{\ell }}{\partial \lambda _{k}}\frac{\partial ^{3}\beta }{\partial \lambda
_{i}\partial \lambda _{j}\partial \lambda _{_{\ell }}}
\nonumber \\
& & \!\!\!\!\!  \!\!\!\!\!  \!\!\!\!\!  \!\!\!\!\! \!\!\!\!\!  \!\!\!\!\!  
+
3\beta _{\ell }\beta
_{k}\frac{\partial \beta _{j}}{\partial \lambda _{k}}\frac{\partial
^{2}\beta _{i}}{\partial \lambda _{j}\partial \lambda _{\ell }}\frac{%
\partial \beta }{\partial \lambda _{i}}
+
4\beta _{i}\beta _{\ell }\frac{%
\partial \beta _{k}}{\partial \lambda _{\ell }}\frac{\partial \beta _{j}}{%
\partial \lambda _{k}}\frac{\partial ^{2}\beta }{\partial \lambda
_{j}\partial \lambda _{i}}
\nonumber \\
& & \!\!\!\!\!  \!\!\!\!\!  \!\!\!\!\!  \!\!\!\!\! \!\!\!\!\!  \!\!\!\!\!  
+4\beta _{i}\beta _{j}\beta _{\ell }\frac{\partial ^{2}\beta _{k}}{\partial
\lambda _{\ell }\partial \lambda _{j}}\frac{\partial ^{2}\beta }{\partial
\lambda _{k}\partial \lambda _{i}}
+
\beta _{\ell }\beta _{k}\beta _{j}\frac{%
\partial ^{3}\beta _{i}}{\partial \lambda _{j}\partial \lambda _{k}\partial
\lambda _{\ell }}\frac{\partial \beta }{\partial \lambda _{i}}
\nonumber \\
& & \!\!\!\!\!  \!\!\!\!\!  \!\!\!\!\!  \!\!\!\!\! \!\!\!\!\!  \!\!\!\!\!  
+
3\beta _{\ell
}\beta _{i}\frac{\partial \beta _{k}}{\partial \lambda _{\ell }}\frac{%
\partial ^{2}\beta }{\partial \lambda _{k}\partial \lambda _{j}}\frac{%
\partial \beta _{j}}{\partial \lambda _{i}}
+
\beta _{\ell }\beta _{k}\frac{%
\partial ^{2}\beta _{j}}{\partial \lambda _{\ell }\lambda _{k}}\frac{%
\partial \beta _{i}}{\partial \lambda _{j}}\frac{\partial \beta }{\partial
\lambda _{i}}
\nonumber \\
& & \!\!\!\!\!  \!\!\!\!\!  \!\!\!\!\!  \!\!\!\!\! \!\!\!\!\!  \!\!\!\!\!  
-
10\beta _{k}\frac{\partial \beta _{j}}{\partial \lambda _{k}}\frac{%
\partial \beta _{i}}{\partial \lambda _{j}}\frac{\partial \beta }{\partial
\lambda _{i}}
-
10\beta _{k}\beta _{j}\frac{\partial ^{2}\beta _{i}}{\partial
\lambda _{j}\partial \lambda _{k}}\frac{\partial \beta }{\partial \lambda
_{i}}
\nonumber \\
& & \!\!\!\!\!  \!\!\!\!\!  \!\!\!\!\!  \!\!\!\!\! \!\!\!\!\!  \!\!\!\!\!  
-
30\beta _{k}\beta _{j}\frac{\partial \beta _{i}}{\partial \lambda _{k}}%
\frac{\partial ^{2}\beta }{\partial \lambda _{i}\partial \lambda _{j}}%
-
10\beta _{i}\beta _{j}\beta _{k}\frac{\partial ^{3}\beta }{\partial \lambda
_{i}\partial \lambda _{j}\partial \lambda _{k}}
\nonumber \\
& & \!\!\!\!\!  \!\!\!\!\!  \!\!\!\!\!  \!\!\!\!\! \!\!\!\!\!  \!\!\!\!\!  
+
35\beta _{k}\frac{\partial \beta _{j}}{\partial \lambda _{k}}\frac{%
\partial \beta }{\partial \lambda _{j}}
+
5\beta _{i}\beta _{j}\frac{\partial
^{2}\beta }{\partial \lambda _{i}\partial \lambda _{j}}
\nonumber \\
& & \!\!\!\!\!  \!\!\!\!\!  \!\!\!\!\!  \!\!\!\!\! \!\!\!\!\!  \!\!\!\!\!  
-
50\beta _{i}\frac{%
\partial \beta }{\partial \lambda _{i}}
+
24\beta
\eea

This leads to the quintic order contribution to the Coleman-Weinberg
potential:
\bea
& = & + \frac{h^{5}}{40\sqrt{2}v}\left( 
\beta + \frac{25}{12}\beta _{i}\frac{d\beta }{d\lambda _{i}}
+
\frac{35}{24}
\beta _{j}\beta _{i}\frac{d^{2}\beta }{d\lambda _{j}d\lambda _{i}}\right.
\nonumber \\
& &
+
\frac{35}{24}\beta _{j}\frac{d\beta _{i}}{d\lambda _{j}}\frac{d\beta }{d\lambda _{i}}
+
\frac{5}{12}\beta _{k}\beta _{j}\beta _{i}\frac{d^{3}\beta }{d\lambda _{k}d\lambda _{j}d\lambda _{i}}
\nonumber \\
& &
+\frac{5}{12}\beta _{k}\frac{d\beta_{j}}{d\lambda _{k}}\frac{d\beta _{i}}{d\lambda _{j}}
\frac{d\beta }{d\lambda_{i}}
+
\frac{5}{12}\beta _{j}\beta _{i}\frac{d^{2}\beta _{i}}{d\lambda_{j}d\lambda _{i}}\frac{d\beta }{d\lambda _{i}}
\nonumber \\
& &
+ \frac{5}{4}\beta_{j}\beta _{k}\frac{d\beta _{i}}{d\lambda _{k}}
\frac{d^{2}\beta }{d\lambda_{j}d\lambda _{i}} 
+
\frac{1}{24}\beta _{i}\beta _{j}\beta _{k}\beta _{\ell }
\frac{\partial^{4}\beta }{\partial \lambda _{i}\partial \lambda _{j}\partial \lambda
_{k}\partial \lambda _{\ell }}
\nonumber \\
& &
+
\frac{1}{24}\beta _{\ell }\frac{\partial
\beta _{k}}{\partial \lambda _{\ell }}\frac{\partial \beta _{j}}{\partial
\lambda _{k}}\frac{\partial \beta _{i}}{\partial \lambda _{j}}\frac{\partial
\beta }{\partial \lambda _{i}}
+
\frac{1}{4}\beta _{i}\beta _{j}\beta _{k}%
\frac{\partial \beta _{\ell }}{\partial \lambda _{k}}\frac{\partial
^{3}\beta }{\partial \lambda _{i}\partial \lambda _{j}\partial \lambda
_{_{\ell }}}
\nonumber \\
& &
+
\frac{1}{8}\beta _{\ell }\beta _{k}\frac{\partial \beta _{j}}{%
\partial \lambda _{k}}\frac{\partial ^{2}\beta _{i}}{\partial \lambda
_{j}\partial \lambda _{\ell }}\frac{\partial \beta }{\partial \lambda _{i}}
+
\frac{1}{6}\beta _{i}\beta _{\ell }\frac{\partial \beta _{k}}{\partial
\lambda _{\ell }}\frac{\partial \beta _{j}}{\partial \lambda _{k}}\frac{%
\partial ^{2}\beta }{\partial \lambda _{j}\partial \lambda _{i}}
\nonumber \\
& &
+
\frac{1}{6}%
\beta _{i}\beta _{j}\beta _{\ell }\frac{\partial ^{2}\beta _{k}}{\partial
\lambda _{\ell }\partial \lambda _{j}}\frac{\partial ^{2}\beta }{\partial
\lambda _{k}\partial \lambda _{i}}
+
\frac{1}{24}\beta _{\ell }\beta _{k}\beta
_{j}\frac{\partial ^{3}\beta _{i}}{\partial \lambda _{j}\partial \lambda
_{k}\partial \lambda _{\ell }}\frac{\partial \beta }{\partial \lambda _{i}}
\nonumber \\
& &
+\left.
\frac{1}{8}\beta _{\ell }\beta _{i}\frac{\partial \beta _{k}}{\partial
\lambda _{\ell }}\frac{\partial ^{2}\beta }{\partial \lambda _{k}\partial
\lambda _{j}}\frac{\partial \beta _{j}}{\partial \lambda _{i}}
+
\frac{1}{24}
\beta _{\ell }\beta _{k}\frac{\partial ^{2}\beta _{j}}{\partial \lambda
_{\ell }\lambda _{k}}\frac{\partial \beta _{i}}{\partial \lambda _{j}}\frac{%
\partial \beta }{\partial \lambda _{i}}\right)
\nonumber \\
& & 
+ O(h^6).
\eea

%%%%%%%%%%%%%%%%%%%%%%%%%%%%%%%%%%%%%%%%%%%%%%%%%%%%%%%%%%%%%%%%%%%%%%%

\end{document}